

%
%
%
%

\catcode `\@=11 

\def\@version{1.6}
\def\@verdate{18th September 1995}

%
%


\newif\ifprod@font

\ifx\@typeface\undefined
  \def\@typeface{Comp. Modern}\prod@fontfalse
\else
  \prod@fonttrue 
\fi

\def\newfam{\alloc@8\fam\chardef\sixt@@n} 

\ifprod@font
\font\fiverm=mtr10 at 5pt
\font\fivebf=mtbx10 at 5pt
\font\fiveit=mtti10 at 5pt
\font\fivesl=mtsl10 at 5pt
\font\fivett=cmtt8 at 5pt     \hyphenchar\fivett=-1
\font\fivecsc=mtcsc10 at 5pt
\font\fivesf=mtss10 at 5pt
\font\fivei=mtmi10 at 5pt      \skewchar\fivei='177
\font\fivesy=mtsy10 at 5pt     \skewchar\fivesy='60

\font\sixrm=mtr10 at 6pt
\font\sixbf=mtbx10 at 6pt
\font\sixit=mtti10 at 6pt
\font\sixsl=mtsl10 at 6pt
\font\sixtt=cmtt8 at 6pt      \hyphenchar\sixtt=-1
\font\sixcsc=mtcsc10 at 6pt
\font\sixsf=mtss10 at 6pt
\font\sixi=mtmi10 at 6pt       \skewchar\sixi='177
\font\sixsy=mtsy10 at 6pt      \skewchar\sixsy='60

\font\sevenrm=mtr10 at 7pt
\font\sevenbf=mtbx10 at 7pt
\font\sevenit=mtti10 at 7pt
\font\sevensl=mtsl10 at 7pt
\font\seventt=cmtt8 at 7pt     \hyphenchar\seventt=-1
\font\sevencsc=mtcsc10 at 7pt
\font\sevensf=mtss10 at 7pt
\font\seveni=mtmi10 at 7pt      \skewchar\seveni='177
\font\sevensy=mtsy10 at 7pt     \skewchar\sevensy='60

\font\eightrm=mtr10 at 8pt
\font\eightbf=mtbx10 at 8pt
\font\eightit=mtti10 at 8pt
\font\eighti=mtmi10 at 8pt      \skewchar\eighti='177
\font\eightsy=mtsy10 at 8pt     \skewchar\eightsy='60
\font\eightsl=mtsl10 at 8pt
\font\eighttt=cmtt8             \hyphenchar\eighttt=-1
\font\eightcsc=mtcsc10 at 8pt
\font\eightsf=mtss10 at 8pt

\font\ninerm=mtr10 at 9pt
\font\ninebf=mtbx10 at 9pt
\font\nineit=mtti10 at 9pt
\font\ninei=mtmi10 at 9pt      \skewchar\ninei='177
\font\ninesy=mtsy10 at 9pt     \skewchar\ninesy='60
\font\ninesl=mtsl10 at 9pt
\font\ninett=cmtt9             \hyphenchar\ninett=-1
\font\ninecsc=mtcsc10 at 9pt
\font\ninesf=mtss10 at 9pt

\font\tenrm=mtr10
\font\tenbf=mtbx10
\font\tenit=mtti10
\font\teni=mtmi10		\skewchar\teni='177
\font\tensy=mtsy10		\skewchar\tensy='60
\font\tenex=cmex10
\font\tensl=mtsl10
\font\tentt=cmtt10		\hyphenchar\tentt=-1
\font\tencsc=mtcsc10
\font\tensf=mtss10

\font\elevenrm=mtr10 at 11pt
\font\elevenbf=mtbx10 at 11pt
\font\elevenit=mtti10 at 11pt
\font\eleveni=mtmi10 at 11pt      \skewchar\eleveni='177
\font\elevensy=mtsy10 at 11pt     \skewchar\elevensy='60
\font\elevensl=mtsl10 at 11pt
\font\eleventt=cmtt10 at 11pt     \hyphenchar\eleventt=-1
\font\elevencsc=mtcsc10 at 11pt
\font\elevensf=mtss10 at 11pt

\font\twelverm=mtr10 at 12pt
\font\twelvebf=mtbx10 at 12pt
\font\twelveit=mtti10 at 12pt
\font\twelvesl=mtsl10 at 12pt
\font\twelvett=cmtt12             \hyphenchar\twelvett=-1
\font\twelvecsc=mtcsc10 at 12pt
\font\twelvesf=mtss10 at 12pt
\font\twelvei=mtmi10 at 12pt      \skewchar\twelvei='177
\font\twelvesy=mtsy10 at 12pt     \skewchar\twelvesy='60

\font\fourteenrm=mtr10 at 14pt
\font\fourteenbf=mtbx10 at 14pt
\font\fourteenit=mtti10 at 14pt
\font\fourteeni=mtmi10 at 14pt      \skewchar\fourteeni='177
\font\fourteensy=mtsy10 at 14pt     \skewchar\fourteensy='60
\font\fourteensl=mtsl10 at 14pt
\font\fourteentt=cmtt12 at 14pt     \hyphenchar\fourteentt=-1
\font\fourteencsc=mtcsc10 at 14pt
\font\fourteensf=mtss10 at 14pt

\font\seventeenrm=mtr10 at 17pt
\font\seventeenbf=mtbx10 at 17pt
\font\seventeenit=mtti10 at 17pt
\font\seventeeni=mtmi10 at 17pt      \skewchar\seventeeni='177
\font\seventeensy=mtsy10 at 17pt     \skewchar\seventeensy='60
\font\seventeensl=mtsl10 at 17pt
\font\seventeentt=cmtt12 at 17pt     \hyphenchar\seventeentt=-1
\font\seventeencsc=mtcsc10 at 17pt
\font\seventeensf=mtss10 at 17pt
\else
\font\fiverm=cmr5
\font\fivei=cmmi5             \skewchar\fivei='177
\font\fivesy=cmsy5            \skewchar\fivesy='60
\font\fivebf=cmbx5

\font\sixrm=cmr6
\font\sixi=cmmi6             \skewchar\sixi='177
\font\sixsy=cmsy6            \skewchar\sixsy='60
\font\sixbf=cmbx6

\font\sevenrm=cmr7
\font\sevenit=cmti7
\font\seveni=cmmi7             \skewchar\seveni='177
\font\sevensy=cmsy7            \skewchar\sevensy='60
\font\sevenbf=cmbx7

\font\eightrm=cmr8
\font\eightbf=cmbx8
\font\eightit=cmti8
\font\eighti=cmmi8			\skewchar\eighti='177
\font\eightsy=cmsy8			\skewchar\eightsy='60
\font\eightsl=cmsl8
\font\eighttt=cmtt8			\hyphenchar\eighttt=-1
\font\eightcsc=cmcsc10 at 8pt
\font\eightsf=cmss8

\font\ninerm=cmr9
\font\ninebf=cmbx9
\font\nineit=cmti9
\font\ninei=cmmi9			\skewchar\ninei='177
\font\ninesy=cmsy9			\skewchar\ninesy='60
\font\ninesl=cmsl9
\font\ninett=cmtt9			\hyphenchar\ninett=-1
\font\ninecsc=cmcsc10 at 9pt
\font\ninesf=cmss9

\font\tenrm=cmr10
\font\tenbf=cmbx10
\font\tenit=cmti10
\font\teni=cmmi10		\skewchar\teni='177
\font\tensy=cmsy10		\skewchar\tensy='60
\font\tenex=cmex10
\font\tensl=cmsl10
\font\tentt=cmtt10		\hyphenchar\tentt=-1
\font\tencsc=cmcsc10
\font\tensf=cmss10

\font\elevenrm=cmr10 scaled \magstephalf
\font\elevenbf=cmbx10 scaled \magstephalf
\font\elevenit=cmti10 scaled \magstephalf
\font\eleveni=cmmi10 scaled \magstephalf	\skewchar\eleveni='177
\font\elevensy=cmsy10 scaled \magstephalf	\skewchar\elevensy='60
\font\elevensl=cmsl10 scaled \magstephalf
\font\eleventt=cmtt10 scaled \magstephalf	\hyphenchar\eleventt=-1
\font\elevencsc=cmcsc10 scaled \magstephalf
\font\elevensf=cmss10 scaled \magstephalf

\font\twelverm=cmr10 scaled \magstep1
\font\twelvebf=cmbx10 scaled \magstep1
\font\twelvei=cmmi10 scaled \magstep1      \skewchar\twelvei='177
\font\twelvesy=cmsy10 scaled \magstep1     \skewchar\twelvesy='60

\font\fourteenrm=cmr10 scaled \magstep2
\font\fourteenbf=cmbx10 scaled \magstep2
\font\fourteenit=cmti10 scaled \magstep2
\font\fourteeni=cmmi10 scaled \magstep2		\skewchar\fourteeni='177
\font\fourteensy=cmsy10 scaled \magstep2	\skewchar\fourteensy='60
\font\fourteensl=cmsl10 scaled \magstep2
\font\fourteentt=cmtt10 scaled \magstep2	\hyphenchar\fourteentt=-1
\font\fourteencsc=cmcsc10 scaled \magstep2
\font\fourteensf=cmss10 scaled \magstep2

\font\seventeenrm=cmr10 scaled \magstep3
\font\seventeenbf=cmbx10 scaled \magstep3
\font\seventeenit=cmti10 scaled \magstep3
\font\seventeeni=cmmi10 scaled \magstep3	\skewchar\seventeeni='177
\font\seventeensy=cmsy10 scaled \magstep3	\skewchar\seventeensy='60
\font\seventeensl=cmsl10 scaled \magstep3
\font\seventeentt=cmtt10 scaled \magstep3	\hyphenchar\seventeentt=-1
\font\seventeencsc=cmcsc10 scaled \magstep3
\font\seventeensf=cmss10 scaled \magstep3
\fi

\def\hexnumber#1{\ifcase#1 0\or1\or2\or3\or4\or5\or6\or7\or8\or9\or
  A\or B\or C\or D\or E\or F\fi}

\def\makestrut{%
  \setbox\strutbox=\hbox{%
    \vrule height.7\baselineskip depth.3\baselineskip width \z@}%
}

\def\baselinestretch{1}
\newskip\tmp@bls

\def\b@ls#1{
  \tmp@bls=#1\relax
  \baselineskip=#1\relax\makestrut
  \normalbaselineskip=\baselinestretch\tmp@bls
  \normalbaselines
}

\def\nostb@ls#1{
  \normalbaselineskip=#1\relax
  \normalbaselines
  \makestrut
}

%

\newfam\scfam  
\newfam\sffam  

\def\mit{\fam\@ne}
\def\cal{\fam\tw@}
\def\em{\ifdim\fontdimen1\font>\z@ \rm\else\it\fi}

\textfont3=\tenex
\scriptfont3=\tenex
\scriptscriptfont3=\tenex

\setbox0=\hbox{\tenex B} \p@renwd=\wd0 

\def\eightpoint{
  \def\rm{\fam0\eightrm}%
  \textfont0=\eightrm \scriptfont0=\sixrm \scriptscriptfont0=\fiverm%
  \textfont1=\eighti  \scriptfont1=\sixi  \scriptscriptfont1=\fivei%
  \textfont2=\eightsy \scriptfont2=\sixsy \scriptscriptfont2=\fivesy%
  \textfont\itfam=\eightit\def\it{\fam\itfam\eightit}%
  \ifprod@font
    \scriptfont\itfam=\sixit
      \scriptscriptfont\itfam=\fiveit
  \else
    \scriptfont\itfam=\eightit
      \scriptscriptfont\itfam=\eightit
  \fi
  \textfont\bffam=\eightbf%
    \scriptfont\bffam=\sixbf%
      \scriptscriptfont\bffam=\fivebf%
  \def\bf{\fam\bffam\eightbf}%
  \textfont\slfam=\eightsl\def\sl{\fam\slfam\eightsl}%
  \ifprod@font
    \scriptfont\slfam=\sixsl
      \scriptscriptfont\slfam=\fivesl
  \else
    \scriptfont\slfam=\eightsl
      \scriptscriptfont\slfam=\eightsl
  \fi
  \textfont\ttfam=\eighttt\def\tt{\fam\ttfam\eighttt}%
  \ifprod@font
    \scriptfont\ttfam=\sixtt
      \scriptscriptfont\ttfam=\fivett
  \else
    \scriptfont\ttfam=\eighttt
      \scriptscriptfont\ttfam=\eighttt
  \fi
  \textfont\scfam=\eightcsc\def\sc{\fam\scfam\eightcsc}%
  \ifprod@font
    \scriptfont\scfam=\sixcsc
      \scriptscriptfont\scfam=\fivecsc
  \else
    \scriptfont\scfam=\eightcsc
      \scriptscriptfont\scfam=\eightcsc
  \fi
  \textfont\sffam=\eightsf\def\sf{\fam\sffam\eightsf}%
  \ifprod@font
    \scriptfont\sffam=\sixsf
      \scriptscriptfont\sffam=\fivesf
  \else
    \scriptfont\sffam=\eightsf
      \scriptscriptfont\sffam=\eightsf
  \fi
  \def\oldstyle{\fam\@ne\eighti}%
  \b@ls{10pt}\rm\@viiipt%
}
\def\@viiipt{}

\def\ninepoint{
  \def\rm{\fam0\ninerm}%
  \textfont0=\ninerm \scriptfont0=\sixrm \scriptscriptfont0=\fiverm%
  \textfont1=\ninei  \scriptfont1=\sixi  \scriptscriptfont1=\fivei%
  \textfont2=\ninesy \scriptfont2=\sixsy \scriptscriptfont2=\fivesy%
  \textfont\itfam=\nineit\def\it{\fam\itfam\nineit}%
  \ifprod@font
    \scriptfont\itfam=\sixit
      \scriptscriptfont\itfam=\fiveit
  \else
    \scriptfont\itfam=\nineit
      \scriptscriptfont\itfam=\nineit
  \fi
  \textfont\bffam=\ninebf%
    \scriptfont\bffam=\sixbf%
      \scriptscriptfont\bffam=\fivebf%
  \def\bf{\fam\bffam\ninebf}%
  \textfont\slfam=\ninesl\def\sl{\fam\slfam\ninesl}%
  \ifprod@font
    \scriptfont\slfam=\sixsl
      \scriptscriptfont\slfam=\fivesl
  \else
    \scriptfont\slfam=\ninesl
      \scriptscriptfont\slfam=\ninesl
  \fi
  \textfont\ttfam=\ninett\def\tt{\fam\ttfam\ninett}%
  \ifprod@font
    \scriptfont\ttfam=\sixtt
      \scriptscriptfont\ttfam=\fivett
  \else
    \scriptfont\ttfam=\ninett
      \scriptscriptfont\ttfam=\ninett
  \fi
  \textfont\scfam=\ninecsc\def\sc{\fam\scfam\ninecsc}%
  \ifprod@font
    \scriptfont\scfam=\sixcsc
      \scriptscriptfont\scfam=\fivecsc
  \else
    \scriptfont\scfam=\ninecsc
      \scriptscriptfont\scfam=\ninecsc
  \fi
  \textfont\sffam=\ninesf\def\sf{\fam\sffam\ninesf}%
  \ifprod@font
    \scriptfont\sffam=\sixsf
      \scriptscriptfont\sffam=\fivesf
  \else
    \scriptfont\sffam=\ninesf
      \scriptscriptfont\sffam=\ninesf
  \fi
  \def\oldstyle{\fam\@ne\ninei}%
  \b@ls{\TextLeading plus \Feathering}\rm\@ixpt%
}
\def\@ixpt{}

\def\tenpoint{
  \def\rm{\fam0\tenrm}%
  \textfont0=\tenrm \scriptfont0=\sevenrm \scriptscriptfont0=\fiverm%
  \textfont1=\teni  \scriptfont1=\seveni  \scriptscriptfont1=\fivei%
  \textfont2=\tensy \scriptfont2=\sevensy \scriptscriptfont2=\fivesy%
  \textfont\itfam=\tenit\def\it{\fam\itfam\tenit}%
  \ifprod@font
    \scriptfont\itfam=\sevenit
      \scriptscriptfont\itfam=\fiveit
  \else
    \scriptfont\itfam=\tenit
      \scriptscriptfont\itfam=\tenit
  \fi
  \textfont\bffam=\tenbf%
    \scriptfont\bffam=\sevenbf%
      \scriptscriptfont\bffam=\fivebf%
  \def\bf{\fam\bffam\tenbf}%
  \textfont\slfam=\tensl\def\sl{\fam\slfam\tensl}%
  \ifprod@font
    \scriptfont\slfam=\sevensl
      \scriptscriptfont\slfam=\fivesl
  \else
    \scriptfont\slfam=\tensl
      \scriptscriptfont\slfam=\tensl
  \fi
  \textfont\ttfam=\tentt\def\tt{\fam\ttfam\tentt}%
  \ifprod@font
    \scriptfont\ttfam=\seventt
      \scriptscriptfont\ttfam=\fivett
  \else
    \scriptfont\ttfam=\tentt
      \scriptscriptfont\ttfam=\tentt
  \fi
  \textfont\scfam=\tencsc\def\sc{\fam\scfam\tencsc}%
  \ifprod@font
    \scriptfont\scfam=\sevencsc
      \scriptscriptfont\scfam=\fivecsc
  \else
    \scriptfont\scfam=\tencsc
      \scriptscriptfont\scfam=\tencsc
  \fi
  \textfont\sffam=\tensf\def\sf{\fam\sffam\tensf}%
  \ifprod@font
    \scriptfont\sffam=\sevensf
      \scriptscriptfont\sffam=\fivesf
  \else
    \scriptfont\sffam=\tensf
      \scriptscriptfont\sffam=\tensf
  \fi
  \def\oldstyle{\fam\@ne\teni}%
  \b@ls{11pt}\rm\@xpt%
}
\def\@xpt{}

\def\elevenpoint{
  \def\rm{\fam0\elevenrm}%
  \textfont0=\elevenrm \scriptfont0=\eightrm \scriptscriptfont0=\sixrm%
  \textfont1=\eleveni  \scriptfont1=\eighti  \scriptscriptfont1=\sixi%
  \textfont2=\elevensy \scriptfont2=\eightsy \scriptscriptfont2=\sixsy%
  \textfont\itfam=\elevenit\def\it{\fam\itfam\elevenit}%
  \ifprod@font
    \scriptfont\itfam=\eightit
      \scriptscriptfont\itfam=\sixit
  \else
    \scriptfont\itfam=\elevenit
      \scriptscriptfont\itfam=\elevenit
  \fi
  \textfont\bffam=\elevenbf%
    \scriptfont\bffam=\eightbf%
      \scriptscriptfont\bffam=\sixbf%
  \def\bf{\fam\bffam\elevenbf}%
  \textfont\slfam=\elevensl\def\sl{\fam\slfam\elevensl}%
  \ifprod@font
    \scriptfont\slfam=\eightsl
      \scriptscriptfont\slfam=\sixsl
  \else
    \scriptfont\slfam=\elevensl
      \scriptscriptfont\slfam=\elevensl
  \fi
  \textfont\ttfam=\eleventt\def\tt{\fam\ttfam\eleventt}%
  \ifprod@font
    \scriptfont\ttfam=\eighttt
      \scriptscriptfont\ttfam=\sixtt
  \else
    \scriptfont\ttfam=\eleventt
      \scriptscriptfont\ttfam=\eleventt
  \fi
  \textfont\scfam=\elevencsc\def\sc{\fam\scfam\elevencsc}%
  \ifprod@font
    \scriptfont\scfam=\eightcsc
      \scriptscriptfont\scfam=\sixcsc
  \else
    \scriptfont\scfam=\elevencsc
      \scriptscriptfont\scfam=\elevencsc
  \fi
  \textfont\sffam=\elevensf\def\sf{\fam\sffam\elevensf}%
  \ifprod@font
    \scriptfont\sffam=\eightsf
      \scriptscriptfont\sffam=\sixsf
  \else
    \scriptfont\sffam=\elevensf
      \scriptscriptfont\sffam=\elevensf
  \fi
  \def\oldstyle{\fam\@ne\eleveni}%
  \b@ls{13pt}\rm\@xipt%
}
\def\@xipt{}

\def\fourteenpoint{
  \def\rm{\fam0\fourteenrm}%
  \textfont0\fourteenrm  \scriptfont0\tenrm  \scriptscriptfont0\sevenrm%
  \textfont1\fourteeni   \scriptfont1\teni   \scriptscriptfont1\seveni%
  \textfont2\fourteensy  \scriptfont2\tensy  \scriptscriptfont2\sevensy%
  \textfont\itfam=\fourteenit\def\it{\fam\itfam\fourteenit}%
  \ifprod@font
    \scriptfont\itfam=\tenit
      \scriptscriptfont\itfam=\sevenit
  \else
    \scriptfont\itfam=\fourteenit
      \scriptscriptfont\itfam=\fourteenit
  \fi
  \textfont\bffam=\fourteenbf%
    \scriptfont\bffam=\tenbf%
      \scriptscriptfont\bffam=\sevenbf%
  \def\bf{\fam\bffam\fourteenbf}%
  \textfont\slfam=\fourteensl\def\sl{\fam\slfam\fourteensl}%
  \ifprod@font
    \scriptfont\slfam=\tensl
      \scriptscriptfont\slfam=\sevensl
  \else
    \scriptfont\slfam=\fourteensl
      \scriptscriptfont\slfam=\fourteensl
  \fi
  \textfont\ttfam=\fourteentt\def\tt{\fam\ttfam\fourteentt}%
  \ifprod@font
    \scriptfont\ttfam=\tentt
      \scriptscriptfont\ttfam=\seventt
  \else
    \scriptfont\ttfam=\fourteentt
      \scriptscriptfont\ttfam=\fourteentt
  \fi
  \textfont\scfam=\fourteencsc\def\sc{\fam\scfam\fourteencsc}%
  \ifprod@font
    \scriptfont\scfam=\tencsc
      \scriptscriptfont\scfam=\sevencsc
  \else
    \scriptfont\scfam=\fourteencsc
      \scriptscriptfont\scfam=\fourteencsc
  \fi
  \textfont\sffam=\fourteensf\def\sf{\fam\sffam\fourteensf}%
  \ifprod@font
    \scriptfont\sffam=\tensf
      \scriptscriptfont\sffam=\sevensf
  \else
    \scriptfont\sffam=\fourteensf
      \scriptscriptfont\sffam=\fourteensf
  \fi
  \def\oldstyle{\fam\@ne\fourteeni}%
  \b@ls{17pt}\rm\@xivpt%
}
\def\@xivpt{}

\def\seventeenpoint{
  \def\rm{\fam0\seventeenrm}%
  \textfont0\seventeenrm  \scriptfont0\twelverm  \scriptscriptfont0\tenrm%
  \textfont1\seventeeni   \scriptfont1\twelvei   \scriptscriptfont1\teni%
  \textfont2\seventeensy  \scriptfont2\twelvesy  \scriptscriptfont2\tensy%
  \textfont\itfam=\seventeenit\def\it{\fam\itfam\seventeenit}%
  \ifprod@font
    \scriptfont\itfam=\twelveit
      \scriptscriptfont\itfam=\tenit
  \else
    \scriptfont\itfam=\seventeenit
      \scriptscriptfont\itfam=\seventeenit
  \fi
  \textfont\bffam=\seventeenbf%
    \scriptfont\bffam=\twelvebf%
      \scriptscriptfont\bffam=\tenbf%
  \def\bf{\fam\bffam\seventeenbf}%
  \textfont\slfam=\seventeensl\def\sl{\fam\slfam\seventeensl}%
  \ifprod@font
    \scriptfont\slfam=\twelvesl
      \scriptscriptfont\slfam=\tensl
  \else
    \scriptfont\slfam=\seventeensl
      \scriptscriptfont\slfam=\seventeensl
  \fi
  \textfont\ttfam=\seventeentt\def\tt{\fam\ttfam\seventeentt}%
  \ifprod@font
    \scriptfont\ttfam=\twelvett
      \scriptscriptfont\ttfam=\tentt
  \else
    \scriptfont\ttfam=\seventeentt
      \scriptscriptfont\ttfam=\seventeentt
  \fi
  \textfont\scfam=\seventeencsc\def\sc{\fam\scfam\seventeencsc}%
  \ifprod@font
    \scriptfont\scfam=\twelvecsc
      \scriptscriptfont\scfam=\tencsc
  \else
    \scriptfont\scfam=\seventeencsc
      \scriptscriptfont\scfam=\seventeencsc
  \fi
  \textfont\sffam=\seventeensf\def\sf{\fam\sffam\seventeensf}%
  \ifprod@font
    \scriptfont\sffam=\twelvesf
      \scriptscriptfont\sffam=\tensf
  \else
    \scriptfont\sffam=\seventeensf
      \scriptscriptfont\sffam=\seventeensf
  \fi
  \def\oldstyle{\fam\@ne\seventeeni}%
  \b@ls{20pt}\rm\@xviipt%
}
\def\@xviipt{}

\lineskip=1pt      \normallineskip=\lineskip
\lineskiplimit=\z@ \normallineskiplimit=\lineskiplimit


\def\,{\relax\ifmmode \mskip\thinmuskip\else \thinspace\fi}
\let\protect=\relax

\long\def\@ifundefined#1#2#3{\expandafter\ifx\csname
  #1\endcsname\relax#2\else#3\fi}




\newtoks\math@groups \math@groups={}
\def\addtom@thgroup#1#2{#1\expandafter{\the#1#2}} 



\def\addtosizeh@ok#1#2#3#4{%
  \expandafter\def\csname @#1pt\endcsname{%
    \def\s@ze{#2}\def\ss@ze{#3}\def\sss@ze{#4}\the\math@groups%
  }%
}



\let\resetsizehook=\addtosizeh@ok


\ifprod@font
  \addtosizeh@ok{viii} {8} {6}  {5}
  \addtosizeh@ok{ix}   {9} {6}  {5}
  \addtosizeh@ok{x}    {10}{7}  {5}
  \addtosizeh@ok{xi}   {11}{8}  {6}
  \addtosizeh@ok{xiv}  {14}{10} {7}
  \addtosizeh@ok{xvii} {17}{12}{10}
\else
  \addtosizeh@ok{viii} {8}     {6}     {5}
  \addtosizeh@ok{ix}   {9}     {6}     {5}
  \addtosizeh@ok{x}    {10}    {7}     {5}
  \addtosizeh@ok{xi}   {10.95} {8}     {6}
  \addtosizeh@ok{xiv}  {14.4}  {10}    {7}
  \addtosizeh@ok{xvii} {17.28} {12}    {10}
\fi

\def\get@font#1#2#3{%
  \edef\fonts@ze{\romannumeral#3}
  \edef\fontn@me{\fonts@ze#1}
  \@ifundefined{\fontn@me}%
    {
     \global\expandafter\font\csname \fontn@me\endcsname=#2 at #3pt}%
    {}%
}

\def\ass@tfont#1#2{%
  \xdef\fam@name{\csname #1\endcsname}%
  \xdef\font@name{\csname #2\endcsname}%
  \let\textfont@name\font@name
  \textfont\fam@name\textfont@name
}

\def\ass@sfont#1#2{%
  \xdef\fam@name{\csname #1\endcsname}%
  \xdef\font@name{\csname #2\endcsname}%
  \let\textfont@name\font@name
  \scriptfont\fam@name\textfont@name
}

\def\ass@ssfont#1#2{%
  \xdef\fam@name{\csname #1\endcsname}%
  \xdef\font@name{\csname #2\endcsname}%
  \let\textfont@name\font@name
  \scriptscriptfont\fam@name\textfont@name
}


\def\NewSymbolFont#1#2{%
  \expandafter\ifx\csname sym#1fam\endcsname\relax 
    \expandafter\newfam\csname sym#1fam\endcsname
    \expandafter\edef\csname sym#1fam\endcsname{\the\allocationnumber}%
    \addtom@thgroup\math@groups{%
      \get@font{#1}{#2}{\s@ze}%
      \ass@tfont{sym#1fam}{\fontn@me}%
      \get@font{#1}{#2}{\ss@ze}%
      \ass@sfont{sym#1fam}{\fontn@me}%
      \get@font{#1}{#2}{\sss@ze}%
      \ass@ssfont{sym#1fam}{\fontn@me}%
    }%
  \else
    \errmessage{Family `#1' already defined}%
  \fi
}


\def\NewMathSymbol#1#2#3#4{%
  \edef\f@mly{\expandafter\hexnumber{\csname sym#3fam\endcsname}}%
  \mathchardef#1="#2\f@mly#4\relax
}


\newif\ifd@f

\def\NewMathDelimiter#1#2#3#4#5#6{%
  \d@ftrue
  \expandafter\ifx\csname sym#3fam\endcsname\relax
    \d@ffalse \errmessage{Family `#3' is not defined}%
  \fi
  \expandafter\ifx\csname sym#5fam\endcsname\relax
    \d@ffalse \errmessage{Family `#5' is not defined}%
  \fi
  \ifd@f
    \edef\f@mly{\expandafter\hexnumber{\csname sym#3fam\endcsname}}%
    \edef\f@mlytw@{\expandafter\hexnumber{\csname sym#5fam\endcsname}}%
    \xdef#1{\delimiter"#2\f@mly #4\f@mlytw@ #6\relax}%
  \fi
}


\def\setboxz@h{\setbox\z@\hbox}
\def\wdz@{\wd\z@}
\def\boxz@{\box\z@}
\def\setbox@ne{\setbox\@ne}
\def\wd@ne{\wd\@ne}

\def\math@atom#1#2{%
   \binrel@{#1}\binrel@@{#2}}
\def\binrel@#1{\setboxz@h{\thinmuskip0mu
  \medmuskip\m@ne mu\thickmuskip\@ne mu$#1\m@th$}%
 \setbox@ne\hbox{\thinmuskip0mu\medmuskip\m@ne mu\thickmuskip
  \@ne mu${}#1{}\m@th$}%
 \setbox\tw@\hbox{\hskip\wd@ne\hskip-\wdz@}}
\def\binrel@@#1{\ifdim\wd2<\z@\mathbin{#1}\else\ifdim\wd\tw@>\z@
 \mathrel{#1}\else{#1}\fi\fi}

\def\m@thit{1}

\def\set@skchar#1{\global\expandafter\skewchar
  \csname\fontn@me\endcsname=#1\relax}

\def\NewMathAlphabet#1#2#3{%
  \def\tst{#3}%
  \ifx\tst\empty\else 
    \expandafter\gdef\csname #1@sc\endcsname{}
  \fi
  \expandafter\def\csname #1\endcsname{
    \protect\csname @#1\endcsname}%
  \expandafter\def\csname @#1\endcsname##1{
    {%
    \begingroup
      \get@font{#1}{#2}{\s@ze}%
      \@ifundefined{#1@sc}{}{\set@skchar{#3}}%
      \ass@tfont{m@thit}{\fontn@me}%
      \get@font{#1}{#2}{\ss@ze}%
      \@ifundefined{#1@sc}{}{\set@skchar{#3}}%
      \ass@sfont{m@thit}{\fontn@me}%
      \get@font{#1}{#2}{\sss@ze}%
      \@ifundefined{#1@sc}{}{\set@skchar{#3}}%
      \ass@ssfont{m@thit}{\fontn@me}%
      \math@atom{##1}{%
      \mathchoice%
        {\hbox{$\m@th\displaystyle##1$}}%
        {\hbox{$\m@th\textstyle##1$}}%
        {\hbox{$\m@th\scriptstyle##1$}}%
        {\hbox{$\m@th\scriptscriptstyle##1$}}}%
    \endgroup
    }%
  }%
}


\newif\iffirstta  \firsttatrue

\def\set@hchar#1{\global\expandafter\hyphenchar
  \csname\fontn@me\endcsname=#1\relax}

\def\NewTextAlphabet#1#2#3{%
  \iffirstta
    \global\firsttafalse
    \newfam\scratchfam
    \edef\scrt@fam{\the\allocationnumber}%
  \fi
  \def\tst{#3}%
  \ifx\tst\empty\else 
    \expandafter\gdef\csname #1@hc\endcsname{}
  \fi
  \expandafter\def\csname #1\endcsname{
    \protect\csname t@#1\endcsname}%
  \long\expandafter\def\csname t@#1\endcsname##1{
    \ifmmode
      \typeout{Warning: do not use \expandafter\string\csname #1\endcsname
        \space in math mode}\fi%
    {%
      \get@font{#1}{#2}{\s@ze}\let\t@xtfnt=\fontn@me\relax
      \@ifundefined{#1@hc}{}{\set@hchar{#3}}%
      \ass@tfont{scrt@fam}{\fontn@me}%
      \get@font{#1}{#2}{\ss@ze}%
      \@ifundefined{#1@hc}{}{\set@hchar{#3}}%
      \ass@sfont{scrt@fam}{\fontn@me}%
      \get@font{#1}{#2}{\sss@ze}%
      \@ifundefined{#1@hc}{}{\set@hchar{#3}}%
      \ass@ssfont{scrt@fam}{\fontn@me}%
      \fam\scratchfam\csname\t@xtfnt\endcsname
    ##1%
    }%
  }%
  \expandafter\def\csname #1shape
    \endcsname{\protect\csname @#1shape\endcsname}%
  \expandafter\def\csname @#1shape\endcsname{
    \ifmmode
      \typeout{Warning: do not use \expandafter\string\csname
        #1shape\endcsname \space in math mode}\fi
      \get@font{#1}{#2}{\s@ze}\let\t@xtfnt=\fontn@me\relax
      \@ifundefined{#1@hc}{}{\set@hchar{#3}}%
      \ass@tfont{scrt@fam}{\fontn@me}%
      \get@font{#1}{#2}{\ss@ze}%
      \@ifundefined{#1@hc}{}{\set@hchar{#3}}%
      \ass@sfont{scrt@fam}{\fontn@me}%
      \get@font{#1}{#2}{\sss@ze}%
      \@ifundefined{#1@hc}{}{\set@hchar{#3}}%
      \ass@ssfont{scrt@fam}{\fontn@me}%
      \fam\scratchfam\csname\t@xtfnt\endcsname
  }%
}


\ifprod@font
  \def\math@itfnt{mtmib10}
  \def\math@syfnt{mtbsy10}
\else
  \def\math@itfnt{cmmib10}
  \def\math@syfnt{cmbsy10}
\fi

\def\m@thsy{2}

\def\bmath{\protect\@bmath}
\def\@bmath#1{%
  {%
  \begingroup
    \get@font{mthit}{\math@itfnt}{\s@ze}\set@skchar{'177}%
    \ass@tfont{m@thit}{\fontn@me}%
    \get@font{mthit}{\math@itfnt}{\ss@ze}\set@skchar{'177}%
    \ass@sfont{m@thit}{\fontn@me}%
    \get@font{mthit}{\math@itfnt}{\sss@ze}\set@skchar{'177}%
    \ass@ssfont{m@thit}{\fontn@me}%
    \get@font{mthsy}{\math@syfnt}{\s@ze}\set@skchar{'60}%
    \ass@tfont{m@thsy}{\fontn@me}%
    \get@font{mthsy}{\math@syfnt}{\ss@ze}\set@skchar{'60}%
    \ass@sfont{m@thsy}{\fontn@me}%
    \get@font{mthsy}{\math@syfnt}{\sss@ze}\set@skchar{'60}%
    \ass@ssfont{m@thsy}{\fontn@me}%
    \math@atom{#1}{%
    \mathchoice%
      {\hbox{$\m@th\displaystyle#1$}}%
      {\hbox{$\m@th\textstyle#1$}}%
      {\hbox{$\m@th\scriptstyle#1$}}%
      {\hbox{$\m@th\scriptscriptstyle#1$}}}%
  \endgroup
  }%
}



\def\diameter{{\ifmmode\mathchoice
{\ooalign{\hfil\hbox{$\displaystyle/$}\hfil\crcr
{\hbox{$\displaystyle\mathchar"20D$}}}}
{\ooalign{\hfil\hbox{$\textstyle/$}\hfil\crcr
{\hbox{$\textstyle\mathchar"20D$}}}}
{\ooalign{\hfil\hbox{$\scriptstyle/$}\hfil\crcr
{\hbox{$\scriptstyle\mathchar"20D$}}}}
{\ooalign{\hfil\hbox{$\scriptscriptstyle/$}\hfil\crcr
{\hbox{$\scriptscriptstyle\mathchar"20D$}}}}
\else{\ooalign{\hfil/\hfil\crcr\mathhexbox20D}}%
\fi}}

\def\sq{\ifmmode\squareforqed\else{\unskip\nobreak\hfil
\penalty50\hskip1em\null\nobreak\hfil\squareforqed
\parfillskip=0pt\finalhyphendemerits=0\endgraf}\fi}
\def\squareforqed{\hbox{\rlap{$\sqcap$}$\sqcup$}}


\def\bbbc{{\mathchoice {\setbox0=\hbox{$\displaystyle\rm C$}\hbox{\hbox
to0pt{\kern0.4\wd0\vrule height0.9\ht0\hss}\box0}}
{\setbox0=\hbox{$\textstyle\rm C$}\hbox{\hbox
to0pt{\kern0.4\wd0\vrule height0.9\ht0\hss}\box0}}
{\setbox0=\hbox{$\scriptstyle\rm C$}\hbox{\hbox
to0pt{\kern0.4\wd0\vrule height0.9\ht0\hss}\box0}}
{\setbox0=\hbox{$\scriptscriptstyle\rm C$}\hbox{\hbox
to0pt{\kern0.4\wd0\vrule height0.9\ht0\hss}\box0}}}}
\def\bbbq{{\mathchoice {\setbox0=\hbox{$\displaystyle\rm
Q$}\hbox{\raise
0.15\ht0\hbox to0pt{\kern0.4\wd0\vrule height0.8\ht0\hss}\box0}}
{\setbox0=\hbox{$\textstyle\rm Q$}\hbox{\raise
0.15\ht0\hbox to0pt{\kern0.4\wd0\vrule height0.8\ht0\hss}\box0}}
{\setbox0=\hbox{$\scriptstyle\rm Q$}\hbox{\raise
0.15\ht0\hbox to0pt{\kern0.4\wd0\vrule height0.7\ht0\hss}\box0}}
{\setbox0=\hbox{$\scriptscriptstyle\rm Q$}\hbox{\raise
0.15\ht0\hbox to0pt{\kern0.4\wd0\vrule height0.7\ht0\hss}\box0}}}}
\def\bbbt{{\mathchoice {\setbox0=\hbox{$\displaystyle\rm
T$}\hbox{\hbox to0pt{\kern0.3\wd0\vrule height0.9\ht0\hss}\box0}}
{\setbox0=\hbox{$\textstyle\rm T$}\hbox{\hbox
to0pt{\kern0.3\wd0\vrule height0.9\ht0\hss}\box0}}
{\setbox0=\hbox{$\scriptstyle\rm T$}\hbox{\hbox
to0pt{\kern0.3\wd0\vrule height0.9\ht0\hss}\box0}}
{\setbox0=\hbox{$\scriptscriptstyle\rm T$}\hbox{\hbox
to0pt{\kern0.3\wd0\vrule height0.9\ht0\hss}\box0}}}}
\def\bbbs{{\mathchoice
{\setbox0=\hbox{$\displaystyle     \rm S$}\hbox{\raise0.5\ht0\hbox
to0pt{\kern0.35\wd0\vrule height0.45\ht0\hss}\hbox
to0pt{\kern0.55\wd0\vrule height0.5\ht0\hss}\box0}}
{\setbox0=\hbox{$\textstyle        \rm S$}\hbox{\raise0.5\ht0\hbox
to0pt{\kern0.35\wd0\vrule height0.45\ht0\hss}\hbox
to0pt{\kern0.55\wd0\vrule height0.5\ht0\hss}\box0}}
{\setbox0=\hbox{$\scriptstyle      \rm S$}\hbox{\raise0.5\ht0\hbox
to0pt{\kern0.35\wd0\vrule height0.45\ht0\hss}\raise0.05\ht0\hbox
to0pt{\kern0.5\wd0\vrule height0.45\ht0\hss}\box0}}
{\setbox0=\hbox{$\scriptscriptstyle\rm S$}\hbox{\raise0.5\ht0\hbox
to0pt{\kern0.4\wd0\vrule height0.45\ht0\hss}\raise0.05\ht0\hbox
to0pt{\kern0.55\wd0\vrule height0.45\ht0\hss}\box0}}}}
\def\bbbz{{\mathchoice {\hbox{$\sf\textstyle Z\kern-0.4em Z$}}
{\hbox{$\sf\textstyle Z\kern-0.4em Z$}}
{\hbox{$\sf\scriptstyle Z\kern-0.3em Z$}}
{\hbox{$\sf\scriptscriptstyle Z\kern-0.2em Z$}}}}


\def\Nulle{0} 
\def\Afe{1}   
\def\Hae{2}   
\def\Hbe{3}   
\def\Hce{4}   
\def\Hde{5}   


\newcount\LastMac       \LastMac=\Nulle

\newskip\half      \half=5.5pt plus 1.5pt minus 2.25pt
\newskip\one       \one=11pt plus 3pt minus 5.5pt
\newskip\onehalf   \onehalf=16.5pt plus 5.5pt minus 8.25pt
\newskip\two       \two=22pt plus 5.5pt minus 11pt

\def\Half{\addvspace{\half}}
\def\One{\addvspace{\one}}
\def\OneHalf{\addvspace{\onehalf}}
\def\Two{\addvspace{\two}}

\def\Raggedright{
  \rightskip=\z@ plus \hsize\relax
}

\def\Fullout{
  \rightskip=\z@\relax
}

\def\Hang#1#2{
  \hangindent=#1%
  \hangafter=#2\relax
}


\newif\ifsp@page
\def\pagestyle#1{\csname ps@#1\endcsname}
\def\thispagestyle#1{\global\sp@pagetrue\gdef\sp@type{#1}}

\def\ps@titlepage{%
  \def\@oddhead{\eightpoint\noindent \the\CatchLine
    \ifprod@font\else\qquad Printed\ \today\qquad
      (MN plain \TeX\ macros\ v\@version)\fi \hfil}%
  \let\@evenhead=\@oddhead
  \def\@oddfoot{\eightpoint\copyright\ \@pubyear\ RAS\hfil}%
  \def\@evenfoot{\hfil\eightpoint\noindent\copyright\ \@pubyear\ RAS}%
}

\def\ps@headings{%
  \def\@oddhead{\elevenpoint\it\noindent
    \hfill\the\RightHeader\hskip1.5em\rm\folio}%
  \def\@evenhead{\elevenpoint\noindent
    \folio\hskip1.5em\it\the\LeftHeader\hfill}%
  \def\@oddfoot{\eightpoint\noindent\copyright\ \@pubyear\ RAS,
    MNRAS {\bf \@volume}, \@pagerange\hfil}%
  \def\@evenfoot{\hfil\eightpoint\copyright\ \@pubyear\ RAS,
    MNRAS {\bf \@volume}, \@pagerange}%
}

\def\ps@plate{%
  \def\@oddhead{\eightpoint\noindent\plt@cap\hfil}%
  \def\@evenhead{\eightpoint\noindent\plt@cap\hfil}%
  \def\@oddfoot{\eightpoint\noindent\copyright\ \@pubyear\ RAS,
    MNRAS {\bf \@volume}, \@pagerange\hfil}%
  \def\@evenfoot{\hfil\eightpoint\copyright\ \@pubyear\ RAS,
    MNRAS {\bf \@volume}, \@pagerange}%
}



\def\title#1{
  \bgroup
    \vbox to 8pt{\vss}%
    \seventeenpoint
    \Raggedright
    \noindent \strut{\bf #1}\par
  \egroup
}

\def\author#1{
  \bgroup
    \ifnum\LastMac=\Afe \OneHalf\else \vskip 21pt\fi
    \fourteenpoint
    \Raggedright
    \noindent \strut #1\par
    \vskip 3pt%
  \egroup
}

\def\affiliation#1{
  \bgroup
    \vskip -4pt%
    \eightpoint
    \Raggedright
    \noindent \strut {\it #1}\par
  \egroup
  \LastMac=\Afe\relax
}

\def\acceptedline#1{
  \bgroup
    \Two
    \eightpoint
    \Raggedright
    \noindent \strut #1\par
  \egroup
}

\long\def\abstract#1{%
  \bgroup
    \vskip 20pt%
    \leftskip 11pc\rightskip\z@
    \noindent{\ninebf ABSTRACT}\par
    \tenpoint
    \Fullout
    \noindent #1\par
  \egroup
}

\long\def\keywords#1{
  \bgroup
    \Half
    \leftskip 11pc\rightskip\z@
    \tenpoint
    \Fullout
    \noindent\hbox{\bf Key words:}\ #1\par
  \egroup
}


\def\maketitle{%
  \EndOpening
  \ifsinglecol \else \MakePage\fi
}


\def\pageoffset#1#2{\hoffset=#1\relax\voffset=#2\relax}


\def\@nameuse#1{\csname #1\endcsname}
\def\arabic#1{\@arabic{\@nameuse{#1}}}
\def\alph#1{\@alph{\@nameuse{#1}}}
\def\Alph#1{\@Alph{\@nameuse{#1}}}
\def\@arabic#1{\number #1}
\def\@Alph#1{\ifcase#1\or A\or B\or C\or D\else\@Ialph{#1}\fi}
\def\@Ialph#1{\ifcase#1\or \or \or \or \or E\or F\or G\or H\or I\or J\or
   K\or L\or M\or N\or O\or P\or Q\or R\or S\or T\or U\or V\or W\or X\or
   Y\or Z\else\errmessage{Counter out of range}\fi}
\def\@alph#1{\ifcase#1\or a\or b\or c\or d\else\@ialph{#1}\fi}
\def\@ialph#1{\ifcase#1\or \or \or \or \or e\or f\or g\or h\or i\or j\or
   k\or l\or m\or n\or o\or p\or q\or r\or s\or t\or u\or v\or w\or x\or y\or
   z\else\errmessage{Counter out of range}\fi}


\newcount\Eqnno
\newcount\SubEqnno

\def\theeq{\arabic{Eqnno}}
\def\thesubeq{\alph{SubEqnno}}

\def\stepeq{\relax
  \global\SubEqnno \z@
  \global\advance\Eqnno \@ne\relax
  {\rm (\theeq)}%
}

\def\startsubeq{\relax
  \global\SubEqnno \z@
  \global\advance\Eqnno \@ne\relax
  \stepsubeq
}

\def\stepsubeq{\relax
  \global\advance\SubEqnno \@ne\relax
  {\rm (\theeq\thesubeq)}%
}


\newcount\Sec        
\newcount\SecSec
\newcount\SecSecSec

\def\thesection{\arabic{Sec}}
\def\thesubsection{\thesection.\arabic{SecSec}}
\def\thesubsubsection{\thesubsection.\arabic{SecSecSec}}

\Sec=\z@

\def\:{\let\@sptoken= } \:  
\def\:{\@xifnch} \expandafter\def\: {\futurelet\@tempc\@ifnch}

\def\@ifnextchar#1#2#3{%
  \let\@tempMACe #1%
  \def\@tempMACa{#2}%
  \def\@tempMACb{#3}%
  \futurelet \@tempMACc\@ifnch%
}

\def\@ifnch{%
\ifx \@tempMACc \@sptoken%
  \let\@tempMACd\@xifnch%
\else%
  \ifx \@tempMACc \@tempMACe%
    \let\@tempMACd\@tempMACa%
  \else%
    \let\@tempMACd\@tempMACb%
  \fi%
\fi%
\@tempMACd%
}

\def\@ifstar#1#2{\@ifnextchar *{\def\@tempMACa*{#1}\@tempMACa}{#2}}

\newskip\@tempskipb

\def\addvspace#1{%
  \ifvmode\else \endgraf\fi%
  \ifdim\lastskip=\z@%
    \vskip #1\relax%
  \else%
    \@tempskipb#1\relax\@xaddvskip%
  \fi%
}

\def\@xaddvskip{%
  \ifdim\lastskip<\@tempskipb%
    \vskip-\lastskip%
    \vskip\@tempskipb\relax%
  \else%
    \ifdim\@tempskipb<\z@%
      \ifdim\lastskip<\z@ \else%
        \advance\@tempskipb\lastskip%
        \vskip-\lastskip\vskip\@tempskipb%
      \fi%
    \fi%
  \fi%
}

\newskip\@tmpSKIP

\def\addpen#1{%
  \ifvmode
    \if@nobreak
    \else
      \ifdim\lastskip=\z@
        \penalty#1\relax
      \else
        \@tmpSKIP=\lastskip
        \vskip -\lastskip
        \penalty#1\vskip\@tmpSKIP
      \fi
    \fi
  \fi
}

\newcount\@clubpen   \@clubpen=\clubpenalty
\newif\if@nobreak    \@nobreakfalse

\def\@noafterindent{%
  \global\@nobreaktrue
  \everypar{\if@nobreak
              \global\@nobreakfalse
              \clubpenalty \@M
              {\setbox\z@\lastbox}%
              \LastMac=\Nulle\relax%
            \else
              \clubpenalty \@clubpen
              \everypar{}%
            \fi}%
}

\newcount\gds@cbrk   \gds@cbrk=-300

\def\@nohdbrk{\interlinepenalty \@M\relax}

\let\@par=\par
\def\@restorepar{\def\par{\@par}}

\newif\if@endpe   \@endpefalse
 
\def\@doendpe{\@endpetrue \@nobreakfalse \LastMac=\Nulle\relax%
     \def\par{\@restorepar\everypar{}\par\@endpefalse}%
              \everypar{\setbox\z@\lastbox\everypar{}\@endpefalse}%
}

\def\section{\@ifstar{\@ssection}{\@section}}

\def\@section#1{
  \if@nobreak
    \everypar{}%
    \ifnum\LastMac=\Hae \addvspace{\half}\fi
  \else
    \addpen{\gds@cbrk}%
    \addvspace{\two}%
  \fi
  \bgroup
    \ninepoint\bf
    \Raggedright
    \global\advance\Sec \@ne
    \ifappendix
      \global\Eqnno=\z@ \global\SubEqnno=\z@\relax
      \def\ch@ck{#1}%
      \ifx\ch@ck\empty \def\c@lon{}\else\def\c@lon{:}\fi
      \noindent\@nohdbrk APPENDIX\ \thesection\c@lon\hskip 0.5em%
        \uppercase{#1}\par
    \else
      \noindent\@nohdbrk\thesection\hskip 1pc \uppercase{#1}\par
    \fi
    \global\SecSec=\z@
  \egroup
  \nobreak
  \vskip\half
  \nobreak
  \@noafterindent
  \LastMac=\Hae\relax
}

\def\@ssection#1{
  \if@nobreak
    \everypar{}%
    \ifnum\LastMac=\Hae \addvspace{\half}\fi
  \else
    \addpen{\gds@cbrk}%
    \addvspace{\two}%
  \fi
  \bgroup
    \ninepoint\bf
    \Raggedright
    \noindent\@nohdbrk\uppercase{#1}\par
  \egroup
  \nobreak
  \vskip\half
  \nobreak
  \@noafterindent
  \LastMac=\Hae\relax
}

\def\subsection{\@ifstar{\@ssubsection}{\@subsection}}

\def\@subsection#1{
  \if@nobreak
    \everypar{}%
    \ifnum\LastMac=\Hae \addvspace{1pt plus 1pt minus .5pt}\fi
  \else
    \addpen{\gds@cbrk}%
    \addvspace{\onehalf}%
  \fi
  \bgroup
    \ninepoint\bf
    \Raggedright
    \global\advance\SecSec \@ne
    \noindent\@nohdbrk\thesubsection \hskip 1pc\relax #1\par
    \global\SecSecSec=\z@
  \egroup
  \nobreak
  \vskip\half
  \nobreak
  \@noafterindent
  \LastMac=\Hbe\relax
}

\def\@ssubsection#1{
  \if@nobreak
    \everypar{}%
    \ifnum\LastMac=\Hae \addvspace{1pt plus 1pt minus .5pt}\fi
  \else
    \addpen{\gds@cbrk}%
    \addvspace{\onehalf}%
  \fi
  \bgroup
    \ninepoint\bf
    \Raggedright
    \noindent\@nohdbrk #1\par
  \egroup
  \nobreak
  \vskip\half
  \nobreak
  \@noafterindent
  \LastMac=\Hbe\relax
}

\def\subsubsection{\@ifstar{\@ssubsubsection}{\@subsubsection}}

\def\@subsubsection#1{
  \if@nobreak
    \everypar{}%
    \ifnum\LastMac=\Hbe \addvspace{1pt plus 1pt minus .5pt}\fi
  \else
    \addpen{\gds@cbrk}%
    \addvspace{\onehalf}%
  \fi
  \bgroup
    \ninepoint\it
    \Raggedright
    \global\advance\SecSecSec \@ne
    \noindent\@nohdbrk\thesubsubsection \hskip 1pc\relax #1\par
  \egroup
  \nobreak
  \vskip\half
  \nobreak
  \@noafterindent
  \LastMac=\Hce\relax
}

\def\@ssubsubsection#1{
  \if@nobreak
    \everypar{}%
    \ifnum\LastMac=\Hbe \addvspace{1pt plus 1pt minus .5pt}\fi
  \else
    \addpen{\gds@cbrk}%
    \addvspace{\onehalf}%
  \fi
  \bgroup
    \ninepoint\it
    \Raggedright
    \noindent\@nohdbrk #1\par
  \egroup
  \nobreak
  \vskip\half
  \nobreak
  \@noafterindent
  \LastMac=\Hce\relax
}

\def\paragraph#1{
  \if@nobreak
    \everypar{}%
  \else
    \addpen{\gds@cbrk}%
    \addvspace{\one}%
  \fi%
  \bgroup%
    \ninepoint\it
    \noindent #1\ \nobreak%
  \egroup
  \LastMac=\Hde\relax
  \ignorespaces
}


\newif\ifappendix

\def\appendix{%
  \global\appendixtrue
  \def\thesection{\Alph{Sec}}%
  \def\thesubsection{\thesection\arabic{SecSec}}%
  \def\theeq{\thesection\arabic{Eqnno}}%
  \Sec=\z@ \SecSec=\z@ \SecSecSec=\z@ \Eqnno=\z@ \SubEqnno=\z@\relax
}




\def\beginlist{%
  \par\if@nobreak \else\addvspace{\half}\fi%
  \bgroup%
    \ninepoint
    \let\item=\list@item%
}

\def\list@item{%
  \par\noindent\hskip 1em\relax%
  \ignorespaces%
}

\def\endlist{\par\egroup\addvspace{\half}\@doendpe}


\def\beginrefs{%
  \par
  \bgroup
    \eightpoint
    \Fullout
    \let\bibitem=\bib@item
}

\def\bib@item{%
  \par\parindent=1.5em\Hang{1.5em}{1}%
  \everypar={\Hang{1.5em}{1}\ignorespaces}%
  \noindent\ignorespaces
}

\def\endrefs{\par\egroup\@doendpe}


\newtoks\CatchLine

\def\@journal{Mon.\ Not.\ R.\ Astron.\ Soc.\ }  
\def\@pubyear{2016}        
\def\@pagerange{000--000}  
\def\@volume{000}          
\def\@microfiche{}         %

\def\pubyear#1{\gdef\@pubyear{#1}\@makecatchline}
\def\pagerange#1{\gdef\@pagerange{#1}\@makecatchline}
\def\volume#1{\gdef\@volume{#1}\@makecatchline}
\def\microfiche#1{\gdef\@microfiche{and Microfiche\ #1}\@makecatchline}

\def\@makecatchline{%
  \global\CatchLine{%
    {\rm \@journal {\bf \@volume},\ \@pagerange\ (\@pubyear)\ \@microfiche}}%
}

\@makecatchline 

\newtoks\LeftHeader
\def\shortauthor#1{
  \global\LeftHeader{#1}%
}

\newtoks\RightHeader
\def\shorttitle#1{
  \global\RightHeader{#1}%
}

\def\PageHead{
  \begingroup
    \ifsp@page
      \csname ps@\sp@type\endcsname
    \fi
    \ifodd\pageno
      \let\the@head=\@oddhead
    \else
      \let\the@head=\@evenhead
    \fi
    \vbox to \z@{\vskip-22.5\p@%
      \hbox to \PageWidth{\vbox to8.5\p@{}%
        \the@head
      }%
    \vss}%
  \endgroup
  \nointerlineskip
}

\gdef\PageFoot{%
  \nointerlineskip%
  \begingroup
  \ifsp@page
    \csname ps@\sp@type\endcsname
    \global\sp@pagefalse
  \fi
  \vbox to 22pt{\vfil%
    \hbox to \PageWidth{%
      \eightpoint\strut\noindent
      \ifodd\pageno
        \@oddfoot
      \else
        \@evenfoot
      \fi
    }%
  }%
  \endgroup
}

\def\today{%
  \number\day\space
  \ifcase\month\or January\or February\or March\or April\or May\or June\or
    July\or August\or September\or October\or November\or December\fi
  \space\number\year%
}

\def\authorcomment#1{%
  \gdef\PageFoot{%
    \nointerlineskip%
    \vbox to 20pt{\vfil%
      \hbox to \PageWidth{\elevenpoint\noindent \hfil #1 \hfil}}%
  }%
}


\newif\ifplate@page
\newbox\plt@box

\def\beginplatepage{%
  \let\plate=\plate@head
  \let\caption=\fig@caption
  \global\setbox\plt@box=\vbox\bgroup
  \TEMPDIMEN=\PageWidth 
  \hsize=\PageWidth\relax
}

\def\endplatepage{\par\egroup\global\plate@pagetrue}
\def\plate@head#1{\gdef\plt@cap{#1}}


\def\letters{%
  \gdef\folio{\ifnum\pageno<\z@ L\romannumeral-\pageno
    \else L\number\pageno \fi}%
}


\newdimen\mathindent

\global\mathindent=\z@
\global\everydisplay{\global\@dspwd=\displaywidth\displaysetup}


\def\@displaylines#1{
  {}$\displ@y\hbox{\vbox{\halign{$\@lign\hfil\displaystyle##\hfil$\crcr
  #1\crcr}}}${}%
}

\def\@eqalign#1{\null\vcenter{\openup\jot\m@th
  \ialign{\strut\hfil$\displaystyle{##}$&$\displaystyle{{}##}$\hfil
      \crcr#1\crcr}}%
}

\def\@eqalignno#1{
  \global\advance\@dspwd by -\mathindent%
  {}$\displ@y\hbox{\vbox{\halign to\@dspwd%
  {\hfil$\@lign\displaystyle{##}$\tabskip\z@skip
  &$\@lign\displaystyle{{}##}$\hfil\tabskip\centering
  &\llap{$\@lign##$}\tabskip\z@skip\crcr
  #1\crcr}}}${}%
}


\global\let\displaylines=\@displaylines
\global\let\eqalign=\@eqalign
\global\let\eqalignno=\@eqalignno
\global\let\leqalignno=\@eqalignno

\newdimen\@dspwd   \@dspwd=\z@
\newif\if@eqno
\newif\if@leqno
\newtoks\@eqn
\newtoks\@eq

\def\displaysetup#1$${\displaytest#1\eqno\eqno\displaytest}

\def\displaytest#1\eqno#2\eqno#3\displaytest{%
 \if!#3!\ldisplaytest#1\leqno\leqno\ldisplaytest
 \else\@eqnotrue\@leqnofalse\@eqn={#2}\@eq={#1}\fi
 \generaldisplay$$}

\def\ldisplaytest#1\leqno#2\leqno#3\ldisplaytest{%
\@eq={#1}%
 \if!#3!\@eqnofalse\else\@eqnotrue\@leqnotrue
  \@eqn={#2}\fi}

\def\generaldisplay{%
  \if@eqno
    \if@leqno
      \hbox to \displaywidth{\noindent
        \rlap{$\displaystyle\the\@eqn$}%
        \hskip\mathindent$\displaystyle\the\@eq$\hfil}%
    \else
      \hbox to \displaywidth{\noindent
        \hskip\mathindent
        $\displaystyle\the\@eq$\hfil$\displaystyle\the\@eqn$}%
    \fi
  \else
    \hbox to \displaywidth{\noindent
      \hskip\mathindent$\displaystyle\the\@eq$\hfil}%
  \fi
}


\def\@notice{%
  \par\Two%
  \noindent{\b@ls{11pt}\ninerm This paper has been produced using the
    Royal Astronomical Society/Blackwell Science \TeX\ macros.\par}%
}

\outer\def\bye{\@notice\par\vfill\supereject\end}


\def\start@mess{%
  Monthly notices of the RAS journal style (\@typeface)\space
    v\@version,\space \@verdate.%
}

\everyjob{\Warn{\start@mess}}



\newif\if@debug \@debugfalse  

\def\Print#1{\if@debug\immediate\write16{#1}\else \fi}
\def\Warn#1{\immediate\write16{#1}}
\def\wlog#1{}

\newcount\Iteration 

\def\Single{0} \def\Double{1}                 
\def\Figure{0} \def\Table{1}                  

\def\InStack{0}  
\def\InZoneA{1}
\def\InZoneB{2}
\def\InZoneC{3}

\newcount\TEMPCOUNT 
\newdimen\TEMPDIMEN 
\newbox\TEMPBOX     
\newbox\VOIDBOX     

\newcount\LengthOfStack 
\newcount\MaxItems      
\newcount\StackPointer
\newcount\Point         
\newcount\NextFigure    
\newcount\NextTable     
\newcount\NextItem      

\newcount\StatusStack   
\newcount\NumStack      
\newcount\TypeStack     
\newcount\SpanStack     
\newcount\BoxStack      

\newcount\ItemSTATUS    
\newcount\ItemNUMBER    
\newcount\ItemTYPE      
\newcount\ItemSPAN      
\newbox\ItemBOX         
\newdimen\ItemSIZE      

\newdimen\PageHeight    
\newdimen\TextLeading   
\newdimen\Feathering    
\newcount\LinesPerPage  
\newdimen\ColumnWidth   
\newdimen\ColumnGap     
\newdimen\PageWidth     
\newdimen\BodgeHeight   
\newcount\Leading       

\newdimen\ZoneBSize  
\newdimen\TextSize   
\newbox\ZoneABOX     
\newbox\ZoneBBOX     
\newbox\ZoneCBOX     

\newif\ifFirstSingleItem
\newif\ifFirstZoneA
\newif\ifMakePageInComplete
\newif\ifMoreFigures \MoreFiguresfalse 
\newif\ifMoreTables  \MoreTablesfalse  

\newif\ifFigInZoneB 
\newif\ifFigInZoneC 
\newif\ifTabInZoneB 
\newif\ifTabInZoneC

\newif\ifZoneAFullPage

\newbox\MidBOX    
\newbox\LeftBOX
\newbox\RightBOX
\newbox\PageBOX   

\newif\ifLeftCOL  
\LeftCOLtrue

\newdimen\ZoneBAdjust

\newcount\ItemFits
\def\Yes{1}
\def\No{2}


\MaxItems=15
\NextFigure=\z@        
\NextTable=\@ne

\BodgeHeight=6pt
\TextLeading=11pt    
\Leading=11
\Feathering=\z@      
\LinesPerPage=61     
\topskip=\TextLeading
\ColumnWidth=20pc    
\ColumnGap=2pc       

\newskip\ItemSepamount  
\ItemSepamount=\TextLeading plus \TextLeading minus 4pt

\parskip=\z@ plus .1pt
\parindent=18pt
\widowpenalty=\z@
\clubpenalty=10000
\tolerance=1500
\hbadness=1500
\abovedisplayskip=6pt plus 2pt minus 1pt
\belowdisplayskip=6pt plus 2pt minus 1pt
\abovedisplayshortskip=6pt plus 2pt minus 1pt
\belowdisplayshortskip=6pt plus 2pt minus 1pt

\frenchspacing

\ninepoint 

\PageHeight=682pt
\PageWidth=2\ColumnWidth
\advance\PageWidth by \ColumnGap

\pagestyle{headings}




\newcount\DUMMY \StatusStack=\allocationnumber
\newcount\DUMMY \newcount\DUMMY \newcount\DUMMY 
\newcount\DUMMY \newcount\DUMMY \newcount\DUMMY 
\newcount\DUMMY \newcount\DUMMY \newcount\DUMMY
\newcount\DUMMY \newcount\DUMMY \newcount\DUMMY 
\newcount\DUMMY \newcount\DUMMY \newcount\DUMMY

\newcount\DUMMY \NumStack=\allocationnumber
\newcount\DUMMY \newcount\DUMMY \newcount\DUMMY 
\newcount\DUMMY \newcount\DUMMY \newcount\DUMMY 
\newcount\DUMMY \newcount\DUMMY \newcount\DUMMY 
\newcount\DUMMY \newcount\DUMMY \newcount\DUMMY 
\newcount\DUMMY \newcount\DUMMY \newcount\DUMMY

\newcount\DUMMY \TypeStack=\allocationnumber
\newcount\DUMMY \newcount\DUMMY \newcount\DUMMY 
\newcount\DUMMY \newcount\DUMMY \newcount\DUMMY 
\newcount\DUMMY \newcount\DUMMY \newcount\DUMMY 
\newcount\DUMMY \newcount\DUMMY \newcount\DUMMY 
\newcount\DUMMY \newcount\DUMMY \newcount\DUMMY

\newcount\DUMMY \SpanStack=\allocationnumber
\newcount\DUMMY \newcount\DUMMY \newcount\DUMMY 
\newcount\DUMMY \newcount\DUMMY \newcount\DUMMY 
\newcount\DUMMY \newcount\DUMMY \newcount\DUMMY 
\newcount\DUMMY \newcount\DUMMY \newcount\DUMMY 
\newcount\DUMMY \newcount\DUMMY \newcount\DUMMY

\newbox\DUMMY   \BoxStack=\allocationnumber
\newbox\DUMMY   \newbox\DUMMY \newbox\DUMMY 
\newbox\DUMMY   \newbox\DUMMY \newbox\DUMMY 
\newbox\DUMMY   \newbox\DUMMY \newbox\DUMMY 
\newbox\DUMMY   \newbox\DUMMY \newbox\DUMMY 
\newbox\DUMMY   \newbox\DUMMY \newbox\DUMMY

\def\wlog{\immediate\write\m@ne}


\def\GetItemAll#1{%
 \GetItemSTATUS{#1}
 \GetItemNUMBER{#1}
 \GetItemTYPE{#1}
 \GetItemSPAN{#1}
 \GetItemBOX{#1}
}

\def\GetItemSTATUS#1{%
 \Point=\StatusStack
 \advance\Point by #1
 \global\ItemSTATUS=\count\Point
}

\def\GetItemNUMBER#1{%
 \Point=\NumStack
 \advance\Point by #1
 \global\ItemNUMBER=\count\Point
}

\def\GetItemTYPE#1{%
 \Point=\TypeStack
 \advance\Point by #1
 \global\ItemTYPE=\count\Point
}

\def\GetItemSPAN#1{%
 \Point\SpanStack
 \advance\Point by #1
 \global\ItemSPAN=\count\Point
}

\def\GetItemBOX#1{%
 \Point=\BoxStack
 \advance\Point by #1
 \global\setbox\ItemBOX=\vbox{\copy\Point}
 \global\ItemSIZE=\ht\ItemBOX
 \global\advance\ItemSIZE by \dp\ItemBOX
 \TEMPCOUNT=\ItemSIZE
 \divide\TEMPCOUNT by \Leading
 \divide\TEMPCOUNT by 65536
 \advance\TEMPCOUNT \@ne
 \ItemSIZE=\TEMPCOUNT pt
 \global\multiply\ItemSIZE by \Leading
}


\def\JoinStack{%
 \ifnum\LengthOfStack=\MaxItems 
  \Warn{WARNING: Stack is full...some items will be lost!}
 \else
  \Point=\StatusStack
  \advance\Point by \LengthOfStack
  \global\count\Point=\ItemSTATUS
  \Point=\NumStack
  \advance\Point by \LengthOfStack
  \global\count\Point=\ItemNUMBER
  \Point=\TypeStack
  \advance\Point by \LengthOfStack
  \global\count\Point=\ItemTYPE
  \Point\SpanStack
  \advance\Point by \LengthOfStack
  \global\count\Point=\ItemSPAN
  \Point=\BoxStack
  \advance\Point by \LengthOfStack
  \global\setbox\Point=\vbox{\copy\ItemBOX}
  \global\advance\LengthOfStack \@ne
  \ifnum\ItemTYPE=\Figure 
   \global\MoreFigurestrue
  \else
   \global\MoreTablestrue
  \fi
 \fi
}


\def\LeaveStack#1{%
 {\Iteration=#1
 \loop
 \ifnum\Iteration<\LengthOfStack
  \advance\Iteration \@ne
  \GetItemSTATUS{\Iteration}
   \advance\Point by \m@ne
   \global\count\Point=\ItemSTATUS
  \GetItemNUMBER{\Iteration}
   \advance\Point by \m@ne
   \global\count\Point=\ItemNUMBER
  \GetItemTYPE{\Iteration}
   \advance\Point by \m@ne
   \global\count\Point=\ItemTYPE
  \GetItemSPAN{\Iteration}
   \advance\Point by \m@ne
   \global\count\Point=\ItemSPAN
  \GetItemBOX{\Iteration}
   \advance\Point by \m@ne
   \global\setbox\Point=\vbox{\copy\ItemBOX}
 \repeat}
 \global\advance\LengthOfStack by \m@ne
}


\newif\ifStackNotClean

\def\CleanStack{%
 \StackNotCleantrue
 {\Iteration=\z@
  \loop
   \ifStackNotClean
    \GetItemSTATUS{\Iteration}
    \ifnum\ItemSTATUS=\InStack
     \advance\Iteration \@ne
     \else
      \LeaveStack{\Iteration}
    \fi
   \ifnum\LengthOfStack<\Iteration
    \StackNotCleanfalse
   \fi
 \repeat}
}


\def\FindItem#1#2{%
 \global\StackPointer=\m@ne 
 {\Iteration=\z@
  \loop
  \ifnum\Iteration<\LengthOfStack
   \GetItemSTATUS{\Iteration}
   \ifnum\ItemSTATUS=\InStack
    \GetItemTYPE{\Iteration}
    \ifnum\ItemTYPE=#1
     \GetItemNUMBER{\Iteration}
     \ifnum\ItemNUMBER=#2
      \global\StackPointer=\Iteration
      \Iteration=\LengthOfStack 
     \fi
    \fi
   \fi
  \advance\Iteration \@ne
 \repeat}
}


\def\FindNext{%
 \global\StackPointer=\m@ne 
 {\Iteration=\z@
  \loop
  \ifnum\Iteration<\LengthOfStack
   \GetItemSTATUS{\Iteration}
   \ifnum\ItemSTATUS=\InStack
    \GetItemTYPE{\Iteration}
   \ifnum\ItemTYPE=\Figure
    \ifMoreFigures
      \global\NextItem=\Figure
      \global\StackPointer=\Iteration
      \Iteration=\LengthOfStack 
    \fi
   \fi
   \ifnum\ItemTYPE=\Table
    \ifMoreTables
      \global\NextItem=\Table
      \global\StackPointer=\Iteration
      \Iteration=\LengthOfStack 
    \fi
   \fi
  \fi
  \advance\Iteration \@ne
 \repeat}
}


\def\ChangeStatus#1#2{%
 \Point=\StatusStack
 \advance\Point by #1
 \global\count\Point=#2
}



\def\Zone{\InZoneA}

\ZoneBAdjust=\z@

\def\MakePage{
 \global\ZoneBSize=\PageHeight
 \global\TextSize=\ZoneBSize
 \global\ZoneAFullPagefalse
 \global\topskip=\TextLeading
 \MakePageInCompletetrue
 \MoreFigurestrue
 \MoreTablestrue
 \FigInZoneBfalse
 \FigInZoneCfalse
 \TabInZoneBfalse
 \TabInZoneCfalse
 \global\FirstSingleItemtrue
 \global\FirstZoneAtrue
 \global\setbox\ZoneABOX=\box\VOIDBOX
 \global\setbox\ZoneBBOX=\box\VOIDBOX
 \global\setbox\ZoneCBOX=\box\VOIDBOX
 \loop
  \ifMakePageInComplete
 \FindNext
 \ifnum\StackPointer=\m@ne
  \NextItem=\m@ne
  \MoreFiguresfalse
  \MoreTablesfalse
 \fi
 \ifnum\NextItem=\Figure
   \FindItem{\Figure}{\NextFigure}
   \ifnum\StackPointer=\m@ne \global\MoreFiguresfalse
   \else
    \GetItemSPAN{\StackPointer}
    \ifnum\ItemSPAN=\Single \def\Zone{\InZoneB}\relax
     \ifFigInZoneC \global\MoreFiguresfalse\fi
    \else
     \def\Zone{\InZoneA}
     \ifFigInZoneB \def\Zone{\InZoneC}\fi
    \fi
   \fi
   \ifMoreFigures\Print{}\FigureItems\fi
 \fi
\ifnum\NextItem=\Table
   \FindItem{\Table}{\NextTable}
   \ifnum\StackPointer=\m@ne \global\MoreTablesfalse
   \else
    \GetItemSPAN{\StackPointer}
    \ifnum\ItemSPAN=\Single\relax
     \ifTabInZoneC \global\MoreTablesfalse\fi
    \else
     \def\Zone{\InZoneA}
     \ifTabInZoneB \def\Zone{\InZoneC}\fi
    \fi
   \fi
   \ifMoreTables\Print{}\TableItems\fi
 \fi
   \MakePageInCompletefalse 
   \ifMoreFigures\MakePageInCompletetrue\fi
   \ifMoreTables\MakePageInCompletetrue\fi
 \repeat
 \ifZoneAFullPage
  \global\TextSize=\z@
  \global\ZoneBSize=\z@
  \global\vsize=\z@\relax
  \global\topskip=\z@\relax
  \vbox to \z@{\vss}
  \eject
 \else
 \global\advance\ZoneBSize by -\ZoneBAdjust
 \global\vsize=\ZoneBSize
 \global\hsize=\ColumnWidth
 \global\ZoneBAdjust=\z@
 \ifdim\TextSize<23pt
 \Warn{}
 \Warn{* Making column fall short: TextSize=\the\TextSize *}
 \vskip-\lastskip\eject\fi
 \fi
}

\def\MakeRightCol{
 \global\TextSize=\ZoneBSize
 \MakePageInCompletetrue
 \MoreFigurestrue
 \MoreTablestrue
 \global\FirstSingleItemtrue
 \global\setbox\ZoneBBOX=\box\VOIDBOX
 \def\Zone{\InZoneB}
 \loop
  \ifMakePageInComplete
 \FindNext
 \ifnum\StackPointer=\m@ne
  \NextItem=\m@ne
  \MoreFiguresfalse
  \MoreTablesfalse
 \fi
 \ifnum\NextItem=\Figure
   \FindItem{\Figure}{\NextFigure}
   \ifnum\StackPointer=\m@ne \MoreFiguresfalse
   \else
    \GetItemSPAN{\StackPointer}
    \ifnum\ItemSPAN=\Double\relax
     \MoreFiguresfalse\fi
   \fi
   \ifMoreFigures\Print{}\FigureItems\fi
 \fi
 \ifnum\NextItem=\Table
   \FindItem{\Table}{\NextTable}
   \ifnum\StackPointer=\m@ne \MoreTablesfalse
   \else
    \GetItemSPAN{\StackPointer}
    \ifnum\ItemSPAN=\Double\relax
     \MoreTablesfalse\fi
   \fi
   \ifMoreTables\Print{}\TableItems\fi
 \fi
   \MakePageInCompletefalse 
   \ifMoreFigures\MakePageInCompletetrue\fi
   \ifMoreTables\MakePageInCompletetrue\fi
 \repeat
 \ifZoneAFullPage
  \global\TextSize=\z@
  \global\ZoneBSize=\z@
  \global\vsize=\z@\relax
  \global\topskip=\z@\relax
  \vbox to \z@{\vss}
  \eject
 \else
 \global\vsize=\ZoneBSize
 \global\hsize=\ColumnWidth
 \ifdim\TextSize<23pt
 \Warn{}
 \Warn{* Making column fall short: TextSize=\the\TextSize *}
 \vskip-\lastskip\eject\fi
\fi
}

\def\FigureItems{
 \Print{Considering...}
 \ShowItem{\StackPointer}
 \GetItemBOX{\StackPointer} 
 \GetItemSPAN{\StackPointer}
  \CheckFitInZone 
  \ifnum\ItemFits=\Yes
   \ifnum\ItemSPAN=\Single
     \ChangeStatus{\StackPointer}{\InZoneB} 
     \global\FigInZoneBtrue
     \ifFirstSingleItem
      \hbox{}\vskip-\BodgeHeight
     \global\advance\ItemSIZE by \TextLeading
     \fi
     \unvbox\ItemBOX\ItemSep
     \global\FirstSingleItemfalse
     \global\advance\TextSize by -\ItemSIZE
     \global\advance\TextSize by -\TextLeading
   \else
    \ifFirstZoneA
     \global\advance\ItemSIZE by \TextLeading
     \global\FirstZoneAfalse\fi
    \global\advance\TextSize by -\ItemSIZE
    \global\advance\TextSize by -\TextLeading
    \global\advance\ZoneBSize by -\ItemSIZE
    \global\advance\ZoneBSize by -\TextLeading
    \ifFigInZoneB\relax
     \else
     \ifdim\TextSize<3\TextLeading
     \global\ZoneAFullPagetrue
     \fi
    \fi
    \ChangeStatus{\StackPointer}{\Zone}
    \ifnum\Zone=\InZoneC \global\FigInZoneCtrue\fi
  \fi
   \Print{TextSize=\the\TextSize}
   \Print{ZoneBSize=\the\ZoneBSize}
  \global\advance\NextFigure \@ne
   \Print{This figure has been placed.}
  \else
   \Print{No space available for this figure...holding over.}
   \Print{}
   \global\MoreFiguresfalse
  \fi
}

\def\TableItems{
 \Print{Considering...}
 \ShowItem{\StackPointer}
 \GetItemBOX{\StackPointer} 
 \GetItemSPAN{\StackPointer}
  \CheckFitInZone 
  \ifnum\ItemFits=\Yes
   \ifnum\ItemSPAN=\Single
    \ChangeStatus{\StackPointer}{\InZoneB}
     \global\TabInZoneBtrue
     \ifFirstSingleItem
      \hbox{}\vskip-\BodgeHeight
     \global\advance\ItemSIZE by \TextLeading
     \fi
     \unvbox\ItemBOX\ItemSep
     \global\FirstSingleItemfalse
     \global\advance\TextSize by -\ItemSIZE
     \global\advance\TextSize by -\TextLeading
   \else
    \ifFirstZoneA
    \global\advance\ItemSIZE by \TextLeading
    \global\FirstZoneAfalse\fi
    \global\advance\TextSize by -\ItemSIZE
    \global\advance\TextSize by -\TextLeading
    \global\advance\ZoneBSize by -\ItemSIZE
    \global\advance\ZoneBSize by -\TextLeading
    \ifFigInZoneB\relax
     \else
     \ifdim\TextSize<3\TextLeading
     \global\ZoneAFullPagetrue
     \fi
    \fi
    \ChangeStatus{\StackPointer}{\Zone}
    \ifnum\Zone=\InZoneC \global\TabInZoneCtrue\fi
   \fi
  \global\advance\NextTable \@ne
   \Print{This table has been placed.}
  \else
  \Print{No space available for this table...holding over.}
   \Print{}
   \global\MoreTablesfalse
  \fi
}


\def\CheckFitInZone{%
{\advance\TextSize by -\ItemSIZE
 \advance\TextSize by -\TextLeading
 \ifFirstSingleItem
  \advance\TextSize by \TextLeading
 \fi
 \ifnum\Zone=\InZoneA\relax
  \else \advance\TextSize by -\ZoneBAdjust
 \fi
 \ifdim\TextSize<3\TextLeading \global\ItemFits=\No
 \else \global\ItemFits=\Yes\fi}
}

\def\BeginOpening{%
  \ninepoint
  \thispagestyle{titlepage}%
  \global\setbox\ItemBOX=\vbox\bgroup%
    \hsize=\PageWidth%
    \hrule height \z@
    \ifsinglecol\vskip 6pt\fi 
}

\let\begintopmatter=\BeginOpening  

\def\EndOpening{%
  \One
  \egroup
  \ifsinglecol
    \box\ItemBOX%
    \vskip\TextLeading plus 2\TextLeading
    \@noafterindent
  \else
    \ItemNUMBER=\z@%
    \ItemTYPE=\Figure
    \ItemSPAN=\Double
    \ItemSTATUS=\InStack
    \JoinStack
  \fi
}


\newif\if@here  \@herefalse

\def\no@float{\global\@heretrue}
\let\nofloat=\relax 

\def\beginfigure{%
  \@ifstar{\global\@dfloattrue \@bfigure}{\global\@dfloatfalse \@bfigure}%
}

\def\@bfigure#1{%
  \par
  \if@dfloat
    \ItemSPAN=\Double
    \TEMPDIMEN=\PageWidth
  \else
    \ItemSPAN=\Single
    \TEMPDIMEN=\ColumnWidth
  \fi
  \ifsinglecol
    \TEMPDIMEN=\PageWidth
  \else
    \ItemSTATUS=\InStack
    \ItemNUMBER=#1%
    \ItemTYPE=\Figure
  \fi
  \bgroup
    \hsize=\TEMPDIMEN
    \global\setbox\ItemBOX=\vbox\bgroup
      \eightpoint\nostb@ls{10pt}%
      \let\caption=\fig@caption
      \ifsinglecol \let\nofloat=\no@float\fi
}

\def\fig@caption#1{%
  \vskip 5.5pt plus 6pt%
  \bgroup 
    \eightpoint\nostb@ls{10pt}%
    \setbox\TEMPBOX=\hbox{#1}%
    \ifdim\wd\TEMPBOX>\TEMPDIMEN
      \noindent \unhbox\TEMPBOX\par
    \else
      \hbox to \hsize{\hfil\unhbox\TEMPBOX\hfil}%
    \fi
  \egroup
}

\def\endfigure{%
  \par\egroup 
  \egroup
  \ifsinglecol
    \if@here \midinsert\global\@herefalse\else \topinsert\fi
      \unvbox\ItemBOX
    \endinsert
  \else
    \JoinStack
    \Print{Processing source for figure \the\ItemNUMBER}%
  \fi
}


\newbox\tab@cap@box
\def\tab@caption#1{\global\setbox\tab@cap@box=\hbox{#1\par}}

\newtoks\tab@txt@toks
\long\def\tab@txt#1{\global\tab@txt@toks={#1}\global\table@txttrue}

\newif\iftable@txt  \table@txtfalse
\newif\if@dfloat    \@dfloatfalse

\def\begintable{%
  \@ifstar{\global\@dfloattrue \@btable}{\global\@dfloatfalse \@btable}%
}

\def\@btable#1{%
  \par
  \if@dfloat
    \ItemSPAN=\Double
    \TEMPDIMEN=\PageWidth
  \else
    \ItemSPAN=\Single
    \TEMPDIMEN=\ColumnWidth
  \fi
  \ifsinglecol
    \TEMPDIMEN=\PageWidth
  \else
    \ItemSTATUS=\InStack
    \ItemNUMBER=#1%
    \ItemTYPE=\Table
  \fi
  \bgroup
    \eightpoint\nostb@ls{10pt}%
    \global\setbox\ItemBOX=\vbox\bgroup
      \let\caption=\tab@caption
      \let\tabletext=\tab@txt
      \ifsinglecol \let\nofloat=\no@float\fi
}

\def\endtable{%
  \par\egroup 
  \egroup
  \setbox\TEMPBOX=\hbox to \TEMPDIMEN{%
    \eightpoint\nostb@ls{10pt}%
    \hss
    \vbox{%
      \hsize=\wd\ItemBOX
      \ifvoid\tab@cap@box
      \else
        \noindent\unhbox\tab@cap@box
        \vskip 5.5pt plus 6pt%
      \fi
      \box\ItemBOX
      \iftable@txt
        \vskip 10pt%
        \noindent\the\tab@txt@toks
        \global\table@txtfalse
      \fi
    }%
    \hss
  }%
  \ifsinglecol
    \if@here \midinsert\global\@herefalse\else \topinsert\fi
      \box\TEMPBOX
    \endinsert
  \else
    \global\setbox\ItemBOX=\box\TEMPBOX
    \JoinStack
    \Print{Processing source for table \the\ItemNUMBER}%
  \fi
}

\def\UnloadZoneA{%
\FirstZoneAtrue
 \Iteration=\z@
  \loop
   \ifnum\Iteration<\LengthOfStack
    \GetItemSTATUS{\Iteration}
    \ifnum\ItemSTATUS=\InZoneA
     \GetItemBOX{\Iteration}
     \ifFirstZoneA \vbox to \BodgeHeight{\vfil}%
     \FirstZoneAfalse\fi
     \unvbox\ItemBOX\ItemSep
     \LeaveStack{\Iteration}
     \else
     \advance\Iteration \@ne
   \fi
 \repeat
}

\def\UnloadZoneC{%
\Iteration=\z@
  \loop
   \ifnum\Iteration<\LengthOfStack
    \GetItemSTATUS{\Iteration}
    \ifnum\ItemSTATUS=\InZoneC
     \GetItemBOX{\Iteration}
     \ItemSep\unvbox\ItemBOX
     \LeaveStack{\Iteration}
     \else
     \advance\Iteration \@ne
   \fi
 \repeat
}


\def\ShowItem#1{
  {\GetItemAll{#1}
  \Print{\the#1:
  {TYPE=\ifnum\ItemTYPE=\Figure Figure\else Table\fi}
  {NUMBER=\the\ItemNUMBER}
  {SPAN=\ifnum\ItemSPAN=\Single Single\else Double\fi}
  {SIZE=\the\ItemSIZE}}}
}

\def\ShowStack{%
 \Print{}
 \Print{LengthOfStack = \the\LengthOfStack}
 \ifnum\LengthOfStack=\z@ \Print{Stack is empty}\fi
 \Iteration=\z@
 \loop
 \ifnum\Iteration<\LengthOfStack
  \ShowItem{\Iteration}
  \advance\Iteration \@ne
 \repeat
}

\def\B#1#2{%
\hbox{\vrule\kern-0.4pt\vbox to #2{%
\hrule width #1\vfill\hrule}\kern-0.4pt\vrule}
}


\newif\ifsinglecol   \singlecolfalse

\def\onecolumn{%
  \global\output={\singlecoloutput}%
  \global\hsize=\PageWidth
  \global\vsize=\PageHeight
  \global\ColumnWidth=\hsize
  \global\TextLeading=12pt
  \global\Leading=12
  \global\singlecoltrue
  \global\let\onecolumn=\relax
  \global\let\footnote=\sing@footnote
  \global\let\vfootnote=\sing@vfootnote
  \ninepoint 
  \message{(Single column)}%
}

\def\singlecoloutput{%
  \shipout\vbox{\PageHead\vbox to \PageHeight{\pagebody\vss}\PageFoot}%
  \advancepageno
  \ifplate@page
    \shipout\vbox{%
      \sp@pagetrue
      \def\sp@type{plate}%
      \global\plate@pagefalse
      \PageHead\vbox to \PageHeight{\unvbox\plt@box\vfil}\PageFoot%
    }%
    \message{[plate]}%
    \advancepageno
  \fi
  \ifnum\outputpenalty>-\@MM \else\dosupereject\fi%
}

\def\ItemSep{\vskip\ItemSepamount\relax}

\def\ItemSepbreak{\par\ifdim\lastskip<\ItemSepamount
  \removelastskip\penalty-200\ItemSep\fi%
}


\let\@@endinsert=\endinsert 

\def\endinsert{\egroup 
  \if@mid \dimen@\ht\z@ \advance\dimen@\dp\z@ \advance\dimen@12\p@
    \advance\dimen@\pagetotal \advance\dimen@-\pageshrink
    \ifdim\dimen@>\pagegoal\@midfalse\p@gefalse\fi\fi
  \if@mid \ItemSep\box\z@\ItemSepbreak
  \else\insert\topins{\penalty100 
    \splittopskip\z@skip
    \splitmaxdepth\maxdimen \floatingpenalty\z@
    \ifp@ge \dimen@\dp\z@
    \vbox to\vsize{\unvbox\z@\kern-\dimen@}
    \else \box\z@\nobreak\ItemSep\fi}\fi\endgroup%
}


\def\gobbleone#1{}
\def\gobbletwo#1#2{}
\let\footnote=\gobbletwo 
\let\vfootnote=\gobbleone

\def\sing@footnote#1{\let\@sf\empty 
  \ifhmode\edef\@sf{\spacefactor\the\spacefactor}\/\fi
  \hbox{$^{\hbox{\eightpoint #1}}$}\@sf\sing@vfootnote{#1}%
}

\def\sing@vfootnote#1{\insert\footins\bgroup\eightpoint\b@ls{9pt}%
  \interlinepenalty\interfootnotelinepenalty
  \splittopskip\ht\strutbox 
  \splitmaxdepth\dp\strutbox \floatingpenalty\@MM
  \leftskip\z@skip \rightskip\z@skip \spaceskip\z@skip \xspaceskip\z@skip
  \noindent $^{\scriptstyle\hbox{#1}}$\hskip 4pt%
    \footstrut\futurelet\next\fo@t%
}

\def\footnoterule{\kern-3\p@ \hrule height \z@ \kern 3\p@}

\skip\footins=19.5pt plus 12pt minus 1pt
\count\footins=1000
\dimen\footins=\maxdimen

\def\note#1#2{%
  \let\@sf=\empty \ifhmode\edef\@sf{\spacefactor\the\spacefactor}\/\fi
  #1\insert\footins\bgroup
    \eightpoint\b@ls{10pt}\rm
    \interlinepenalty\interfootnotelinepenalty
    \splitmaxdepth\dp\strutbox \floatingpenalty\@MM
    \leftskip\z@skip \rightskip\z@skip \spaceskip\z@skip \xspaceskip\z@skip
    \noindent\footstrut #1$\,$#2\strut\par
  \egroup
  \@sf\relax}


\def\landscape{%
  \global\TEMPDIMEN=\PageWidth
  \global\PageWidth=\PageHeight
  \global\PageHeight=\TEMPDIMEN
  \global\let\landscape=\relax
  \onecolumn
  \message{(landscape)}%
  \raggedbottom
}


\output{%
  \ifLeftCOL
    \global\setbox\LeftBOX=\vbox to \ZoneBSize{\box255\unvbox\ZoneBBOX
      \ifvoid\footins\else
        \vskip\skip\footins\unvbox\footins\fi
    }%
    \global\LeftCOLfalse
    \MakeRightCol
  \else
    \setbox\RightBOX=\vbox to \ZoneBSize{\box255\unvbox\ZoneBBOX
      \ifvoid\footins\else
        \vskip\skip\footins\unvbox\footins\fi
    }%
    \setbox\MidBOX=\hbox{\box\LeftBOX\hskip\ColumnGap\box\RightBOX}%
    \setbox\PageBOX=\vbox to \PageHeight{%
      \UnloadZoneA\box\MidBOX\UnloadZoneC}%
    \shipout\vbox{\PageHead\vbox to \PageHeight{\box\PageBOX\vss}\PageFoot}%
    \advancepageno
    \ifplate@page
      \shipout\vbox{%
        \sp@pagetrue
        \def\sp@type{plate}%
        \global\plate@pagefalse
        \PageHead\vbox to \PageHeight{\unvbox\plt@box\vfil}\PageFoot%
      }%
      \message{[plate]}%
      \advancepageno
    \fi
    \global\LeftCOLtrue
    \CleanStack
    \MakePage
  \fi
}


\Warn{\start@mess}

\newif\ifCUPmtplainloaded 
\ifprod@font
  \global\CUPmtplainloadedtrue
\fi


\catcode `\@=12 



\input epsf


\let\sec=\section
\let\ssec=\subsection


 at 11truept

\def\japref{\parskip =0pt\par\noindent\hangindent\parindent
    \parskip =2ex plus .5ex minus .1ex}
\def\gs{\mathrel{\lower0.6ex\hbox{$\buildrel {\textstyle >}
 \over {\scriptstyle \sim}$}}}
\def\ls{\mathrel{\lower0.6ex\hbox{$\buildrel {\textstyle <}
 \over {\scriptstyle \sim}$}}}
\newcount\equationo
\equationo = 0

\def\leftdisplay#1$${\leftline{$\displaystyle{#1}$
  \global\advance\equationo by1\hfill (\the\equationo )}$$}
\everydisplay{\leftdisplay}

\def\m@th{\mathsurround=0pt }
\def\eqaligntwo#1{\null\,\vcenter{\openup1\jot \m@th
 \ialign{\strut$\displaystyle{##}$\hfil&$\displaystyle{##}$\hfil&$\displaystyle{{}##}$\hfil
 \crcr#1\crcr}}\,}

\def\japfigsky#1#2#3{
\beginfigure*{#1}
\epsfxsize=0.87\hsize
\centerline{\epsfbox[72 288 524 554]{#2}}
\caption{%
{\bf Figure #1.}
#3
}
\endfigure
}

\def\japfig#1#2#3#4{
\ifnum #2 = 1
\beginfigure{#1}
\epsfxsize=8.2cm
\centerline{\epsfbox{#3}}
\fi
\ifnum #2 = 2
\beginfigure*{#1}
\epsfxsize=0.88\hsize
\centerline{\epsfbox{#3}}
\fi
\ifnum #2 = 3 
\beginfigure{#1}
\epsfxsize=8.2cm
\centerline{\epsfbox[60 208 510 588]{#3}}
\fi
\ifnum #2 = 4 
\beginfigure{#1}
\epsfxsize=8.2cm
\centerline{\epsfbox[53 15 465 785]{#3}}
\fi
\ifnum #2 = 5 
\beginfigure{#1}
\epsfxsize=6.5cm
\centerline{\epsfbox{#3}}
\fi
\caption{%
{\bf Figure #1.}
#4
}
\endfigure
}



%

\pageoffset{-0.8cm}{-0.3cm}




\overfullrule 0pt

\begintopmatter  

\vglue-2.2truecm
\centerline{\strut}
\vglue 1.7truecm

\title{The SuperCOSMOS all-sky galaxy catalogue}

\author{J.A. Peacock${}^1$, N.C. Hambly${}^1$, M. Bilicki${}^2$, H.T. MacGillivray${}^1$, \hfill\break 
L. Miller${}^3$, M.A. Read${}^1$ and S.B. Tritton${}^1$}
\smallskip
\affiliation{%
${}^1$ Institute for Astronomy, University of Edinburgh, 
Royal Observatory, Blackford Hill, Edinburgh EH9 3HJ \hfill\break
${}^2$ Sterrewacht Leiden, Universiteit Leiden, Niels Bohrweg 2, 2333 CA Leiden, Netherlands \hfill\break
${}^3$ Department of Astrophysics, University of Oxford,
Denys Wilkinson Building, Keble Road, Oxford OX1 3RH
}

\shortauthor{J.A. Peacock et al.}

\shorttitle{The SuperCOSMOS all-sky galaxy catalogue}


\abstract{%
We describe the construction of an all-sky galaxy catalogue, using
SuperCOSMOS scans of Schmidt photographic plates from the UKST and
POSS2 surveys.  The photographic photometry is calibrated
using SDSS data, with results that are linear to 2\%
or better. All-sky photometric uniformity is achieved by
matching plate overlaps and also by requiring homogeneity in optical-to-2MASS
colours, yielding zero
points that are uniform to 0.03 mag. or better.  The typical AB depths
achieved are $B_{\rm J}<21$, $R_{\rm F}<19.5$ and $I_{\rm N}<18.5$, with little difference
between hemispheres. In practice, the $I_{\rm N}$
plates are shallower than the $B_{\rm J}$ \& $R_{\rm F}$ plates, so for most purposes
we advocate the use of a catalogue selected in these two latter bands.  At
high Galactic latitudes, this catalogue is approximately 90\% complete
with 5\% stellar contamination; we quantify how the quality
degrades towards the Galactic plane. At low latitudes, there are many
spurious galaxy candidates resulting from stellar blends:
these approximately match the surface density of true galaxies
at $|b|=30^\circ$. Above this latitude, the catalogue limited in $B_{\rm J}$ \& $R_{\rm F}$
contains in total about 20 million galaxy candidates, of which
75\% are real. This contamination can be removed, and the sky coverage
extended, by matching with additional datasets.
This SuperCOSMOS catalogue has been matched with 2MASS
and with WISE, yielding quasi-allsky samples of respectively 1.5 million and 18.5
million galaxies, to median redshifts of 0.08 and 0.20.
This legacy dataset thus continues to offer a valuable
resource for large-angle cosmological investigations.
}

\keywords{methods: observational; techniques: photometric; catalogues; surveys; galaxies: photometry}

\maketitle  

\sec{INTRODUCTION}

Large galaxy catalogues are an essential tool for any cosmological
study that aims to inspect the large-scale distribution of
matter in the universe. Although the microwave background gives
a purer probe of cosmological deviations from homogeneity, the
pattern of galaxy clustering is the most direct and spectacular
manifestation of these density fluctuations. Historically,
statistical studies of this clustering have contributed hugely
to the establishment of the current flat vacuum-dominated
cosmological standard model (e.g. Efstathiou, Sutherland \& Maddox 1990;
Efstathiou et al. 2002), which has since been confirmed in many different ways (e.g. 
Astier \& Pain 2012; Aubourg et al. 2015; Planck Collaboration XIII 2015).

The older generation of work in this area was dominated by
photographic plates; for many years, digital
detectors were incapable of surveying the required areas of sky.
The most influential photographic catalogue was the APM survey
(Maddox et al. 1990a,b), which was based on scans of the
southern-hemisphere UK Schmidt blue survey plates. This contained
about 20 million galaxies over 4300~$\rm deg^2$ to blue magnitudes
as faint as 22, and formed the original input catalogue for
the highly successful 2dF Galaxy Redshift Survey
(Colless et al. 2001, 2003), 
and (in part) for the 6dF Galaxy survey (Jones et al. 2009). 
Newer generations of optical imaging surveys
are however based on CCD data, and the 5-band Sloan Digital
Sky Survey sets the standard in this respect, having released data
for 208 million galaxies over 31,637~$\rm deg^2$ of imaging, of
which over 1\% have spectroscopic redshifts
(DR12: Alam et al. 2015).

Despite their technological obsolescence, the legacy 
photographic Schmidt surveys nevertheless retain one key advantage: they
cover the whole sky. Digital surveys 
have already reached this stage in the infrared (2MASS: Skrutskie et al. 2006; WISE: Wright et al. 2010) 
or ultraviolet (GALEX: Morrissey et al. 2007), but this is not yet the case in optical wavebands. 
For the latter we will undoubtedly achieve this goal
in due course; but in the meantime there are a variety of
science applications that require such data over the
full sky, and which can achieve interesting results using
the existing material -- whose quality turns out to be perhaps higher
than hitherto suspected, in a tribute to its creators.
The SuperCOSMOS measuring machine was therefore used
to scan the best available photographic data and extract the
full information in the plates. This process was initially carried out
in the southern hemisphere, based on the UK Schmidt Survey
(UKST; Hambly et al. 2001a,b,c); to this has been added scans
of the second-epoch Palomar Survey (POSS2; Reid et al. 1991). 
The photographic material, hypersensitization strategy, filter set
and overall sensitivity of UKST and POSS2 are broadly comparable,
allowing a reasonably homogeneous coverage of the sky.
The photometric
calibration of both these surveys in a uniform and consistent
manner is possible, thanks to the SDSS and in particular also to
the all-sky coverage offered in the 1-2 micron wavebands
by the Two-Micron All-Sky Survey (2MASS;
e.g. Jarrett et al. 2000, 2003). Although it may
seem implausible that near-IR data could be used to calibrate optical
data, it turns out that the statistical power of insisting
on uniform optical-to-IR colours greatly aids the robustness of the
calibration.

This calibration was carried out in 2007; the resulting 
publicly available SuperCOSMOS Science Archive
has since been curated by Edinburgh's Wide-Field Astronomy Unit
({\tt surveys.roe.ac.uk/ssa}). The galaxy catalogue
has been used in past work involving one of the present authors
(Francis \& Peacock 2010a,b). It was also employed to generate
photometric redshifts for the 2MASS galaxy catalogue
(2MPZ: Bilicki et al. 2014), yielding 1.5 million galaxies to a median
redshift of 0.08, with a typical redshift precision of $\sigma_z\simeq 0.015$.
More recently, we have extended this exercise to produce
a joint optical-WISE allsky galaxy catalogue (Bilicki et al. 2016);
this contains 18.5 million galaxies to a median redshift of
0.20, with a typical redshift precision of $\sigma_z/(1+z) \simeq 0.033$.
In addition, several other surveys have benefited from the 
SuperCOSMOS data, including HIPASS (Doyle et al. 2005), 6dFGS (Jones et al. 2009) or AT20G (Murphy et al. 2010).

In order to document the material used in the above work, and to assist other users of
the public data, this paper describes the SuperCOSMOS calibration process in some detail.
Section 2 describes the input SuperCOSMOS data; Section 3 gives colour
equations to SDSS and discusses variations of photometry with position
within a given plate; Section 4 discusses the achievement of all-sky
uniformity with the aid of 2MASS; Section 5 discusses the completeness
and reliability of the catalogue; and Section 6 sums up.

\sec{SUPERCOSMOS DATA}

The operation of the SuperCOSMOS plate measuring machine and its
application to the production of object catalogues is described by
Hambly et al. (2001a,b,c). The machine itself has since been decommissioned,
following the completion of its scanning programme.
In brief, plates were digitized at relatively
high spatial resolution (0.67 arcsec pixels) and high dynamic range (15 bit). 
The photographic transmission values were measured by a scanning CCD 
and converted to an estimate of
linear intensity. Image analysis was then carried out by grouping connected pixels
that lie above a threshold, from which basic image parameters could be deduced,
particularly isophotal magnitude and image area. 

Morphological classification of SuperCOSMOS image data is in essence
based on the area of images as a function of magnitude. Each
image is allocated a profile statistic, $\eta$, which is a measure
of image sharpness scaled to have
zero mean and unit standard deviation
(see Hambly et al. 2001b for the details of
how this is defined). Images for which $\eta$ lies above 2.5 are
classified as extended sources -- i.e. galaxy candidates.
This classification can be performed separately for each
plate, but an improved overall classification can be derived
when the catalogues from different plates are paired up.
Where a given object is detected in more than one waveband,
the $\eta$ values are summed; the image is classified as extended
if $\sum_i\eta_i/\sqrt{N_{\rm detections}}>2.5$
(SuperCOSMOS parameter $\tt meanClass=1$).
Where an image is classified as extended, an attempt is made to use the
brightness profile within the above-threshold pixels to extrapolate
a correction of the isophotal photometry to an approximate total magnitude, assuming
that the brightness profile can be fitted by a Gaussian.

This method is more likely to yield clean samples of stars, since merged pairs of
stars will be classed as galaxies. 
The SuperCOSMOS image analyser does attempt to deblend images, based on
data at multiple thresholds. This is relatively successful for brighter
objects, but inevitably the bulk of objects lie near the
plate limit and thus have sufficient S/N for detection as extended sources, but not sufficient to
diagnose image blends. Various unpublished attempts have been made
over the years to use more sophisticated image classification algorithms,
but without a great deal of success. As a result, a significant problem
with galaxy catalogues obtained from these photographic plates is corruption
by merged stellar images at low Galactic latitudes. This is also a
potential issue with digital data, of course, but the modest image
quality of the Schmidt plates (${\rm FWHM} \simeq 2''$) exacerbates the problem
-- although the difficulty is greater still with the $6''$ FWHM of WISE
(Bilicki et al. 2016; Kurcz et al. 2016).

In addition to morphological classification, each image is assigned a quality flag designed
to pick out problematic images that are likely to be plate artefacts or corrupted in
some way. This information is available for each plate ({\tt qualB, qualR2, qualI}) -- where
`R2' distinguishes the default red data from scans of the shallower plates from the ESO Schmidt
and the first-generation Palomar Schmidt (`R1').
We reject objects where either {\tt qualB} or {\tt qualR2} $\ge 2048$, indicating that the image 
probably results from a stellar halo or diffraction spike, or a satellite trail.

Catalogues of stellar and galaxy images generated in this way have been
produced for photographic plates from the UKST and POSS2
Schmidt surveys. Together, these provide approximately homogeneous
all-sky coverage in three wavebands: the IIIa-J, IIIa-F and IV-N plates,
which we shall generally refer to here as $B_{\rm J}$, $R_{\rm F}$, and $I_{\rm N}$ surveys. This
information has been available on-line since 2001 at
{\tt www-wfau.roe.ac.uk/sss}; the data are now also held in a relational database at
{\tt surveys.roe.ac.uk/ssa}, which provides multicolour information
on 1.9 billion distinct objects.
It should be emphasised that
the calibration of SuperCOSMOS photometry for objects classed as stellar
was carried out separately (see Hambly et al. 2001b), and the
current exercise was applied only to the extended images. Although
it is possible to assign galaxy-style magnitudes to SuperCOSMOS
objects that are not classed as extended by going back to the raw data,
this was not done for the public dataset. Thus stars and galaxies in
the SSA have magnitudes that refer to two slightly different photometric systems.

\ssec{Plate special cases}

In principle, both UKST and POSS2 consist of 894 fields, starting with
number 1 at the south and north celestial poles, respectively. In practice,
field 1 was never taken for POSS2, and the four fields 895--898 were obtained
instead to bracket the pole.

It should be noted that, in one case, it was not possible to obtain a blue
POSS2 plate for scanning; field 504 in the northern hemisphere was therefore
taken with the UKST, and the blue data for this field refer to that system.
Similarly, for the POSS2 $I_{\rm N}$ survey, 25 fields were taken
by UKST to complete the survey:
559, 560, 565, 567, 569, 630, 633, 636, 637, 641, 688, 701, 703, 704, 708, 709, 713,
771, 776, 777, 779, 780, 782, 784, 796.

Both surveys were carried out using grids of field centres spaced by $5^\circ$ in 
B1950 declination, with the $\delta=0$ band being observed by both telescopes.
We took the decision to use UKST data in this case, so that POSS2 measurements are only
present for $\delta_{\rm B1950}>2.5^\circ$. Because of this survey grid, it should
only be necessary to use the data in the central $5^\circ \times 5^\circ$ of
the $6^\circ \times 6^\circ$ Schmidt plates. In practice, where multiple
plates cover a given sky location, the plate where the data lie closest to the 
plate centre is adopted in the final catalogue.
But the full area was scanned and stored,
which had two benefits. Firstly, the overlap aids in the calibration of
relative plate zero-points, as discussed below. Secondly, the data in the
outer parts of the plates permits a cure for the `stepwedge problem'. The 
stepwedges were added to the plates by exposing standard lamps in an
attempt to aid calibration. For UKST, these affect the data only outside
$5^\circ \times 5^\circ$, but for POSS2 the effect comes closer to
the plate centre. The location of the stepwedge is known and masked out,
but this has the result that a small fraction of the sky covered by POSS2
contains no galaxies in 
publicly available merged source catalogues. We have remedied this by
returning to the original scans and using data outside the central
$5^\circ \times 5^\circ$ from overlapping plates (about 2\% of the
northern hemisphere was affected).

\japfig{1}{2}{scos_coleq.eps}
{The colour relations with respect to SDSS photometry for the various
SuperCOSMOS photographic bands, using limits of $B_{\rm J}<20$,
$R_{\rm F}<19.5$, $I_{\rm N}<19$.}

\sec{CALIBRATION WITH SDSS}

\ssec{Colour equations}

Photographic magnitudes have traditionally been placed on
a Vega system and calibrated with respect to 
Johnson--Cousins magnitudes (see e.g. Blair \& Gilmore 1982).
Today, we should consider this in the context of the
largest dataset of calibrating photometry, which is
the SDSS; here
the magnitudes are on an AB system (Fukugita et al. 1996).
In practice, we used the SDSS Petrosian magnitudes as the
`truth' for each galaxy.
Before the photographic magnitudes can be calibrated against such digital
photometry, they must be corrected to a genuine logarithmic measure
of flux -- a non-trivial process, given the highly nonlinear response
of photographic emulsions. The initial SuperCOSMOS processing attempted
to linearise the plate transmission data, but this is hard to
achieve precisely, since saturation effects can vary within the
scanning resolution. As a result, the SuperCOSMOS photometry is
not precisely linear. 
This issue is discussed below in Section 3.3; it must be approached in
an iterative way, since the colour equations discussed here require
linearised data, but plate calibration requires the colour equations
to be known. In practice, an initial plate calibration was made assuming
published colour equations (discussed below), following which direct
colour equations to SDSS could be deduced, allowing a new calibration;
the colour equations were stable thereafter.
The relation between SDSS and SuperCOSMOS magnitudes is significantly nonlinear, requiring
a transformation that is quadratic in SDSS colour (see Fig. 1).

The question of the zero point requires some care.
The primary photometric standard for the SDSS system is BD+17$^\circ$4708,
which has the colours $g-r=0.29$ and $r-i=0.1$; we therefore chose to
set the SuperCOSMOS zero point so that SDSS and
SuperCOSMOS magnitudes are identical for galaxies with
$g-r=0.29$ (in the case of the $B_{\rm J}$ and $R_{\rm F}$ magnitudes)
or $r-i=0.1$ (in the case of $I_{\rm N}$ magnitudes). This means
that our magnitudes are not precisely on an AB system, since this
would require the magnitudes to match at $g-r=0$ and $r-i=0$.
Very few galaxies are this blue, and
we prefer to avoid an extrapolation away from the bulk of 
the calibrating data.
But in any case, Fig. 1 suggests that any correction to
exact AB would be small.

With this choice, the defining equations for
the final SuperCOSMOS magnitudes are:
$$
\eqaligntwo{
B_{\rm J}&= g -0.024\Delta_{g-r} +0.144\Delta_{g-r}^2 &\quad (\rm North) \cr
B_{\rm J}&= g -0.015\Delta_{g-r} +0.102\Delta_{g-r}^2 &\quad (\rm South) \cr
R_{\rm F}&= r -0.133\Delta_{g-r} +0.112\Delta_{g-r}^2 &\quad (\rm North) \cr
R_{\rm F}&= r -0.085\Delta_{g-r} +0.074\Delta_{g-r}^2 &\quad (\rm South) \cr
I_{\rm N}&= i -0.081\Delta_{r-i} +0.254\Delta_{r-i}^2 &\quad (\rm North) \cr
I_{\rm N}&= i -0.151\Delta_{r-i} +0.236\Delta_{r-i}^2 &\quad (\rm South) \cr
}
$$
where $\Delta_{g-r}\equiv (g-r)-0.29$ and $\Delta_{r-i}\equiv (r-i)-0.1$.
But for some purposes, it may be convenient to have the best-fitting linear relations, which
are nearly as accurate:
$$
\eqaligntwo{
B_{\rm J}&= g -0.078 +0.134(g-r)  &\quad (\rm North) \cr
B_{\rm J}&= g -0.058 +0.102(g-r)  &\quad (\rm South) \cr
R_{\rm F}&= r +0.012 -0.054(g-r)  &\quad (\rm North) \cr
R_{\rm F}&= r -0.002 -0.022(g-r)  &\quad (\rm South) \cr
I_{\rm N}&= i +0.008 -0.024(r-i)  &\quad (\rm North) \cr
I_{\rm N}&= i +0.022 -0.092(r-i)  &\quad (\rm South) \cr
}
$$
These relations are intended to apply for $0\ls g-r \ls 1.5$
and $0.1 \ls r-i \ls 0.7$. The rms accuracy for the 
quadratic fit is around 0.005 mag. in $B_{\rm J}$ and $R_{\rm F}$
and 0.015 mag. in $I_{\rm N}$;
for the linear fit, the precision is a factor 1.3 poorer.

For relation to older work, it may be useful to express these
relations in terms of the Johnson--Cousins \hbox{\it UBVRI\/} system.
Colour equations have been given that relate the SDSS
data to the Johnson--Cousins system (Fukugita et al. 1996):
$$
\eqalign{
B&=g+0.217+0.419(g-r) \cr
V&=g-0.002-0.513(g-r) \cr
R&=r-0.155-0.089(g-r) \cr
I&=i-0.389-0.020(r-i)-0.089(g-r)\cr
}
$$
(the last relation is only claimed to apply for $r-i<0.9$, but this should
apply for almost all galaxies here).
The Fukugita et al. results were based on synthetic photometry,
and so are vulnerable to uncertainties in filter profiles. A
safer route is to use direct comparisons of photometry. This has
been performed by Ivezi\'c et al. (2007), comparing Johnson--Cousins
\hbox{\it UBVRI\/} photometry from Stetson (2000, 2005) directly to SDSS magnitudes.
Stetson's photometry is based on extensive CCD observations
of the standards defined by Landolt (1992) using photomultipliers.
Systematics in the relation of the two systems are at the 0.001 mag.
level, so that Stetson's photometry may be regarded as defining
the Johnson--Cousins system in practice. 
The colour equations to SDSS are found to be nonlinear, and
Ivezi\'c et al. give a set of transformations that involve a
cubic in colour. But again, a set of linear transformations may be more
convenient, and the most useful form is an unpublished
set of relations due to Lupton (2005):
$$
\eqalign{
  B &= g + 0.313(g - r) + 0.227 \cr
  V &= g - 0.578(g - r) - 0.004 \cr
  R &= r - 0.184(g - r) - 0.097 \cr
  R &= r - 0.294(r - i) - 0.144 \cr
  I &= i - 0.244(r - i) - 0.382 \cr
}
$$
We adopt these as the primary definitions of Johnson-Cousins magnitudes;
in all cases, the scatter around these relations is close to 0.01 mag. In this
case, and using our linear relations between SuperCOSMOS and SDSS, the
SuperCOSMOS magnitudes may be expressed in terms of \hbox{\it BVRI\/} as
$$
\eqaligntwo{
B_{\rm J}&= B -0.259 -0.201(B-V) &\quad (\rm North) \cr 
B_{\rm J}&= B -0.230 -0.237(B-V) &\quad (\rm South) \cr 
R_{\rm F}&= R +0.089 +0.215(V-R) &\quad (\rm North) \cr 
R_{\rm F}&= R +0.070 +0.267(V-R) &\quad (\rm South) \cr 
I_{\rm N}&= I +0.335 +0.231(R-I) &\quad (\rm North) \cr 
I_{\rm N}&= I +0.356 +0.160(R-I) &\quad (\rm South) \cr 
}
$$

\ssec{Hemispheric colour-dependent offsets}

From the above equations, it can be seen that some slight differences
exist in the POSS2 and UKST photometric systems, despite the use of
closely comparable observing materials. Ideally, one would
like to eliminate these offsets. The most basic approach
would be to adopt a typical colour and deduce an offset in
zero point. For objects that are not detected in all bands, there
may be no other option. From Fig. 1, $g-r=0.8$ and $r-i=0.4$ are
reasonable choices, yielding
$$
\eqalign{
B_{\rm J}^{\rm North} - B_{\rm J}^{\rm South} &= +0.006 \cr
R_{\rm F}^{\rm North} - R_{\rm F}^{\rm South} &= -0.015 \cr
I_{\rm N}^{\rm North} - I_{\rm N}^{\rm South} &= +0.023 \cr
}
$$
In themselves, these offsets are insignificant. But if colours
are available, one can do better. 
Because the $I_{\rm N}$ plates
are less deep, this equalisation exercise is mainly of interest
for the $B_{\rm J}$ and $R_{\rm F}$ magnitudes -- 
since the deepest and most reliable catalogue arises from
requiring detections in both of the shortest bands. In this case,
one can use the following direct relations:
$$
\eqalign{
B_{\rm J}^{\rm North} - B_{\rm J}^{\rm South} &= 
0.03 (B-R)^2 
- 0.005 (B-R)   \cr
R_{\rm F}^{\rm North} - R_{\rm F}^{\rm South} &= 
0.03 (B-R)^2 - 0.06 (B-R) + 0.015 \; .\cr
}
$$
Here, $B-R$ denotes either photographic colour: the $N-S$ difference
is unimportant in making this correction. 
For those applications where the $I_{\rm N}$-plate depth is sufficient (e.g. 2MPZ: Bilicki et al. 2014) 
the corresponding correction is as follows:
$$
I_{\rm N}^{\rm North} - I_{\rm N}^{\rm South} =  0.023 (R-I)^2 + 0.06(R-I) - 0.01 \; .
$$
Note that the North-South corrections given in Bilicki et al. (2014) were incorrect,
owing to an inadvertent swapping of hemispheres in their derivation.

\japfig{2}{2}{trends_north.ps}
{The derived calibration parameters for the POSS2 photographic plates, showing
linearity and zero point as a function of time (plate number), together with
the running-average trends that were used as an initial estimate of the calibration
when no SDSS photometry was available.}

\japfig{3}{2}{trends_south.ps}
{The southern equivalent of Fig. 2: derived calibration parameters for the UKST photographic plates, showing
linearity and zero point as a function of time (plate number), together with
the running-average trends that were used as an initial estimate of the calibration
when no SDSS photometry was available.}

\japfig{4}{2}{scos_magresid.eps}
{Magnitude residuals relative to SDSS, showing the uniformity and linearity of the SuperCOSMOS
photometry after correction of individual plates. The principal selection for
calibrating galaxies was $B_{\rm J}<20$.}

\japfig{5}{1}{ripple_all.ps}
{The effective flat-field correction required 
to account for additional sensitivity variations in the
SuperCOSMOS CCD.}

\japfig{6}{2}{2mpz.ps}
{Histograms of extinction-corrected optical--2MASS colours after 
adjustment of plate zero points. The average colour on each plate is
used to yield an estimate of the zero-point for that plate, as well
as limits on the radially-dependent field effects.}

\ssec{Linearity and zero-points}

Although SuperCOSMOS attempted to correct the measured photographic transmission
to linear intensity, this is hard to carry out precisely. Each plate
has its individual response to hypersensitization, and the finite scan
resolution inevitably blends in saturated data from the centre of images.
The issue of linearity therefore has to be treated empirically. For each
plate, we fitted the following model:
$$
m_{\rm true} = a + b(m_{\rm raw}+2.5).
$$
The offset in the second term is arbitrary, but was chosen to
correspond to the typical SuperCOSMOS instrumental magnitudes
for objects of reasonable $S/N$ that were
chosen for calibration -- in order to remove any strong
statistical coupling between the `zero point', $a$, and the
`linearity', $b$.

The resulting calibration parameters are shown in Figs 2 \& 3, plotted
against the progress of the surveys in time, via the unique number of
each plate (as opposed to the field number). Some significant trends
are visible, most notably the blue zero point, which responds to secular
variations in the brightness of the night sky. The redder zero points
and the linearity show smaller drifts with time. Generally the linearity
is unity within 5-10\%, and after correction the photometry may be assumed
to be accurately linear -- as may be verified by the residual
plots shown in Fig. 4.

For plates where direct calibration is not available,
the trends in Figs 2 \& 3 yield an estimate of the calibration.
The precision seems to be a little better for UKST than for POSS2:
an rms of 2\% (POSS2) / 1\% (UKST) in linearity and 
0.16 mag. (POSS2) / 0.10 mag. (UKST) in zero point. For plates
where direct calibration is lacking, this linearity estimate was
adopted; for the zero point, one can do better -- as described below.

\ssec{Corrections for vignetting and other field effects}

The SDSS information also allows us to test for uniformity of the
photometry within a given plate. It was expected that such variations would
exist, if for no other reason than that geometrical vignetting within
Schmidt telescopes results in less light reaching the edges of plates.
We find that the mean offset in magnitude as a function of radius
can be accurately described by
$$
\Delta m = { 0.095(r-2.25^\circ)^2 \over 1 + 0.025(r-2.25^\circ)^2 },
$$
and zero for $r<2.25^\circ$. However, this behaviour is not universal,
and individual plates exhibit radial trends that differ from this
behaviour. After some experimentation, the following expression was
found to be capable of capturing the empirical effect:
$$
\Delta m = s\, (r-1.5^\circ)^2,
$$
and zero for $r<1.5^\circ$. The `slump coefficient' $s$ needs to be determined
for each plate; typical values are a few hundredths of a magnitude (generally positive,
corresponding to loss of sensitivity at large $r$).
The origin of such `field effects' has been attributed
to differential plate desensitisation being affected by a varying gap between
a flat filter and a curved focal surface. Whether or not this is the correct
explanation, it does seem that the main effect is radially symmetric.

The only exception to the last statement is that small-scale variations
were found perpendicular to the scan direction. These repeated over a distance
corresponding to the SuperCOSMOS CCD and thus are plausibly attributable to
uncorrected flat-field errors in the CCD. The mean effect is easily measured
and is shown in Fig. 5. The correction
is small enough that it was treated as being universal for all plates and
wavebands.

\ssec{Galactic extinction}

For many science applications, we will want to select galaxies to a limit that is
uniform with respect to magnitudes corrected for foreground extinction. This
can be estimated using the colour equations given above. In order
to use tabulated extinction coefficients for SDSS $ugriz$, we use the
above linearised colour equations, expressed as a
weighted linear combinations of magnitudes. These weights 
are then applied to the SDSS
coefficients given in the 
$R_V=3.1$ column from Table 6 of Schlafly \& Finkbeiner (2011).
Note that this recalibration tends to yield coefficients that are
substantially smaller than those given by
Schlegel, Finkbeiner \& Davis (1998), in part because the 
$E(B-V)$ estimates given by the latter authors are generally too high by about 16\%.
Schlafly \& Finkbeiner's coefficients absorb this shift, and are
designed to work with the published $E(B-V)$ estimates. We follow
this convention for SuperCOSMOS, where the 
coefficients in $A=a\, E(B-V)_{\rm SFD}$ are
$3.303,2.285,1.698$ for $(g,r,i)$, yielding
$$
\eqalign{
a(B_{\rm J})&=3.44 \quad (\rm North) \cr
a(B_{\rm J})&=3.41 \quad (\rm South) \cr
a(R_{\rm F})&=2.23 \quad (\rm North) \cr
a(R_{\rm F})&=2.26 \quad (\rm South) \cr
a(I_{\rm N})&=1.68 \quad (\rm North) \cr
a(I_{\rm N})&=1.64 \quad (\rm South) \cr
}
$$

\sec{CALIBRATION WITH 2MASS}

The SDSS calibration is essential for establishing the general systematic
properties of the photographic data, but the whole point of the current exercise is
that this calibration is unavailable over most of the sky. Although the general
properties of plates seem predictable from their position in the observing
sequence, this is not sufficient to tie down all properties
(especially the zero points) to an interesting precision.

The approach taken by the APM team was to bootstrap the calibration of plates
using the 1.25-degree overlap between plates. Although we incorporate this information,
a more robust solution is possible through the near-infrared
2MASS Extended Source Catalogue (XSC; Jarrett et al. 2000; 2003).
We used the centroid coordinates ({\tt sup\_ra} and {\tt sup\_dec}) 
to find SuperCOSMOS counterparts of XSC sources and adopted
the isophotal 2MASS photometry ({\tt j\_m\_k20fe} etc.). These magnitudes may have
small systematic aperture corrections with redshift, but they are the most
stable and therefore suitable for calibration via colour.
The exact 2MASS XSC completeness limit is slightly soft. The design 
completeness was $K<13.5$, but the number counts rise steeply
until $K=14$. We adopt a limit 0.5 mag. brighter than the internal
catalogue limit of 14.3: $K<13.8$ (extinction-corrected). We also
require corrected $J<15.0$.

\japfig{7}{1}{magerrs.ps}
{Magnitude errors as a function of depth, using a clipped rms. Solid points are UKST;
open are POSS2. The different bands are $B_{\rm J}$ (blue); $R_{\rm F}$ (red); $I_{\rm N}$ (black).}

One can then look at the extinction-corrected 
optical-to IR colours of galaxies on plates where SDSS
calibration exists. Histograms of these colours are shown in Fig. 6. The
breadth of these distributions naturally depends on wavelength, declining
from a FWHM in colour of about 0.8 for $B_{\rm J}-J$ to a FWHM of 0.35 for 
$I_{\rm N}-J$. With $\sim 1000$ 2MASS galaxies per plate, that means that the
average colour on each plate can be determined to a formal accuracy of 0.01
or better.
We adjust the optical zero point on a given plate to match the average
optical$-$2MASS colour to the value known from plates with direct optical
calibration. This should then in principle yield optical zero points that 
are uniform to the all-sky photometric precision of 2MASS calibration,
which is claimed to be 0.03 magnitudes.

An earlier version of this procedure was used for refining the plate-dependent
calibration of the APM magnitudes used by the 2dFGRS.
But in addition to giving an estimate of the zero point for each plate,
we found in the present work that
the number of 2MASS matches per plate is sufficient to permit an
estimate of the slump coefficient, by determining the average colour
in radial bins and fitting to equation (10). This means that relative
offsets in plate zero points can be determined for adjoining plates by
matching galaxies in the overlap regions. The zero points were then
iterated to improve agreement between plates and their neighbours,
allowing a maximum shift of 0.05 magnitudes, in order to remain
consistent with the expected absolute precision that should be
delivered by the initial zero points estimated from the 2MASS
matching.

The plate zero points used here were derived
and applied to the SuperCOSMOS ssa database 
at a time when the best available
SDSS dataset was DR6 (Adelman-McCarthy et al. 2008).
The subsequent expansion
of the SDSS would therefore now allow direct calibration for some plates where
the zero point was previously only inferred indirectly as above, but we
have not updated the calibration in these cases. Rather, this paper serves
as a documentation of the public SuperCOSMOS dataset in the form
in which it has been available and used for a number of years.

\sec{RELIABILITY OF THE CATALOGUE}

In addition to calibration of the magnitudes, SDSS 
galaxy catalogues (treated as perfect for the current purposes)
allow us to assess the limitations of the photographic data.

\ssec{Trends with magnitude}

Initially, we restrict attention to high Galactic latitudes
($|b|>60^\circ$), in order to explore the limits of
the data in the case where Galactic 
extinction and contamination is minimised.
Fig. 7 shows the (sigma-clipped) dispersion in magnitude between SuperCOSMOS
data and SDSS calibration, going deeper than the calibration
constraints shown in Fig. 4. Effective magnitude limits are documented
in Table 1. These are rather conservative, since Fig. 7 shows that
the magnitude errors are never less than about 0.1 mag., even well above
any plausible plate limit. This reflects a variety of limitations of the
SuperCOSMOS data: emulsion measurement noise and the finite dynamic range
of the SuperCOSMOS imaging system;
imperfections in the photographic
aperture corrections; and scatter in the colour relations with respect to SDSS.
Thus a fairer estimate of the random noise in the SuperCOSMOS magnitudes
would probably subtract 0.1 in quadrature.
In the face of these figures, there is a level of freedom in choosing limits 
to define a catalogue for practical use. We generally adopt the
limits $B_{\rm J}<21$, $R_{\rm F}<19.5$: this takes the blue plates to
their maximum depth, while requiring in addition a
somewhat more precise detection in the red. This latter criterion permits 
more accurate colours, and is also effective at removing a range of
artefacts that would exist in a single-plate catalogue
(see Fig. 10 below).

\bigskip

\vbox{
\centerline{{\bf Table 1}: SuperCOSMOS magnitude limits (see Fig. 7).}
\bigskip
\centerline{\vbox to 0.5\hsize{
\tabskip 1em
\halign{#\hfill & \hfill#\hfill & \hfill#\hfill \cr
\noalign{\hrule}
\noalign{\vglue 0.2em}
\noalign{\hrule}
\noalign{\vglue 0.5em}
Band & $5\sigma$ & $4\sigma$ \cr
\noalign{\vglue 0.5em}
\noalign{\hrule}
\noalign{\vglue 0.3em}
UKST $B_{\rm J}$ & 20.79 & 21.19 \cr
UKST $R_{\rm F}$ & 19.95 & 20.30 \cr
UKST $I_{\rm N}$ & 18.56 & 19.94 \cr
POSS2 $B_{\rm J}$ & 20.26 & 21.17 \cr
POSS2 $R_{\rm F}$ & 19.78 & 20.35 \cr
POSS2 $I_{\rm N}$ & 18.38 & 18.90 \cr
\noalign{\vglue 0.3em}
\noalign{\hrule}
\noalign{\vglue 0.2em}
\noalign{\hrule}
}}}
}

\ssec{Fidelity}

The remaining question is how well the SuperCOSMOS catalogue approximates
an ideal galaxy catalogue, subject to the magnitude errors shown
in Fig. 7. We need to assess the purity and completeness of the
catalogue: what fraction of our objects are actually galaxies, and
what fraction of true galaxies are catalogued?
This issue is discussed at some length in Section 4.4 of Bilicki et al. (2016),
with the conclusion that at high Galactic latitudes SuperCOSMOS
attains a completeness of about 90\%, with 5\% stellar contamination. 
Inevitably, given the poorer image quality, compact galaxies are
systematically lost from the sample and classified as stars.
But this problem also afflicts digital surveys: in 2MASS, 
for instance, there is growing evidence that about half of the detected
galaxies are in its {\it Point\/} Source Catalogue rather than the list
of extended objects (Kov{\'a}cs \& Szapudi 2015; Rahman et al. 2016).
Even SDSS misses about 2\% of galaxies in this way (Baldry et
al. 2010).

\japfig{8}{1}{density.ps}
{The total surface density of an extinction-corrected sample
cut at $B_{\rm J}<21$ and $R_{\rm F}<19.5$, as a function of Galactic
latitude, for various bins of Galactic longitude.
With the exception of glitches arising from the Magellanic clouds, the surface density
of galaxy images is relatively constant for $|b|\gs 30^\circ$, but stellar blend
artefacts increase the apparent density by more than an order of magnitude at lower latitudes.}

For many purposes, the incompleteness and contamination
figures documented above are tolerable. However, things 
degenerate at low Galactic latitudes.
Fig. 8 shows the surface density as a function of
sky position, where the main visible effect is a
level of contamination rising to low Galactic latitudes
and within the Magellanic Clouds, 
as stellar blending becomes more frequent. At
the galactic poles, the surface density asymptotes to 
about $800\,{\rm deg}^{-2}$ for $B_{\rm J}<21$, $R_{\rm F}<19.5$,
implying in principle 33 million real galaxies over the whole sky.
But the density of artefacts rises steeply at low latitudes,
typically doubling the apparent density at $|b|=30^\circ$; at
$|b|=15^\circ$, the raw surface density is approximately
10 times higher than at the Galactic poles.
Thus the catalogue only has reasonable reliability over
about half the sky (approximately 20 million galaxy 
candidates at $|b|>30^\circ$, excluding the Magellanic clouds,
of which about 75\% are real).

Attempts were made to identify such spurious images
based on discrepant extinction-corrected colours,
but these were rather unsuccessful: the overlap of stellar
and galaxy colours near the plate limits is too great.
A significant reduction of this contamination is
only possible when pairing with other wide-angle
datasets. By using WISE, it was possible to generate
a catalogue of 18.5 million galaxies with low
contamination over 70\% of the sky (Bilicki et al. 2016).

\japfigsky{9}{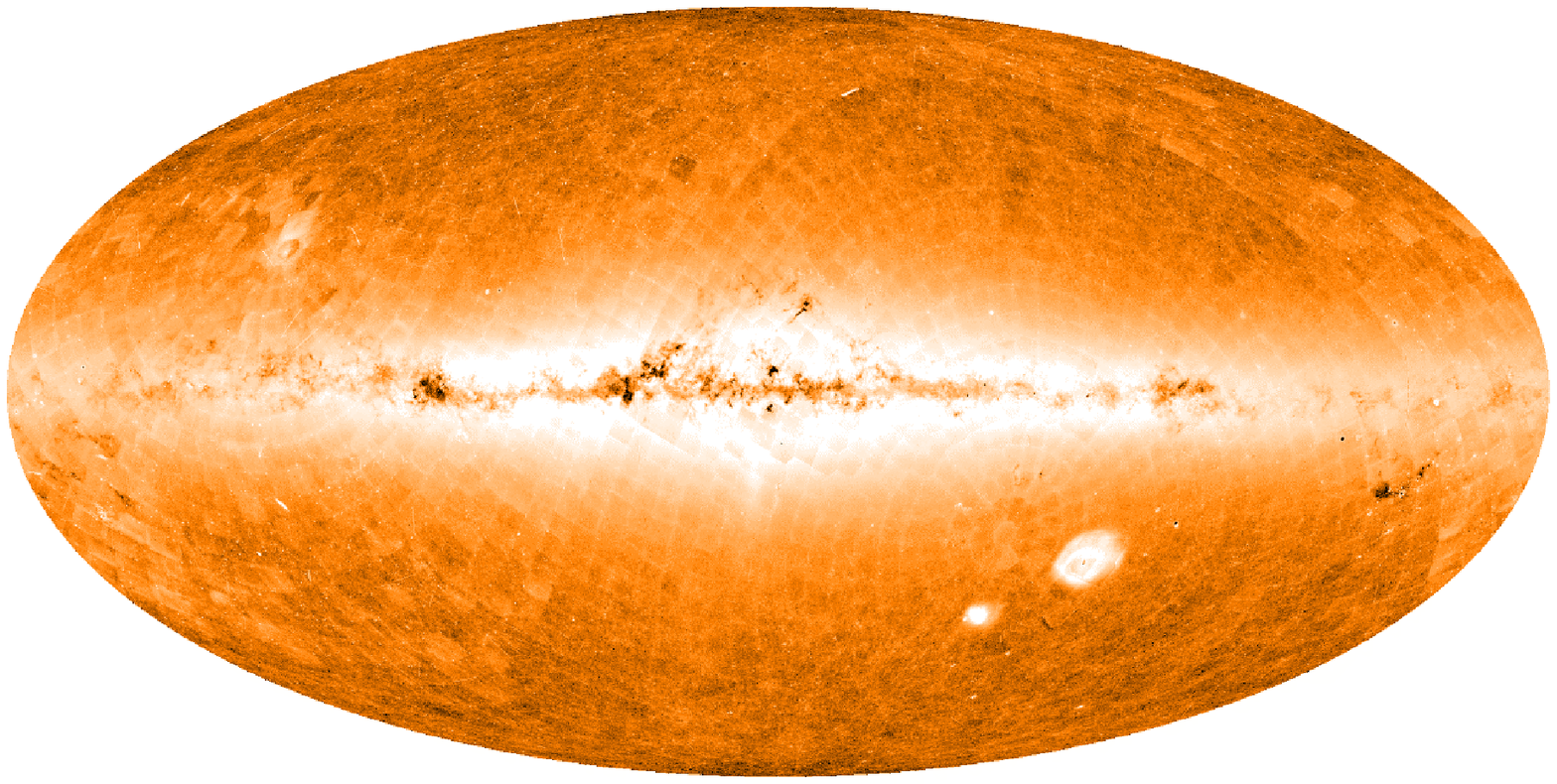}
{The sky distribution of the allsky catalogue, to an
extinction-corrected limit of $B_{\rm J}<21$, without a constraint
on other wavebands. The Mollweide projection in Galactic coordinates
shows surface density on a logarithmic scale, from $0.1\,{\rm arcmin}^{-2}$
to $10\,{\rm arcmin}^{-2}$. The hemispheric offset between the UKST and POSS2 photometric
systems is apparent, and a number of artefacts can be seen. The latter are exposed
in more detail in the next image.
}

\japfigsky{10}{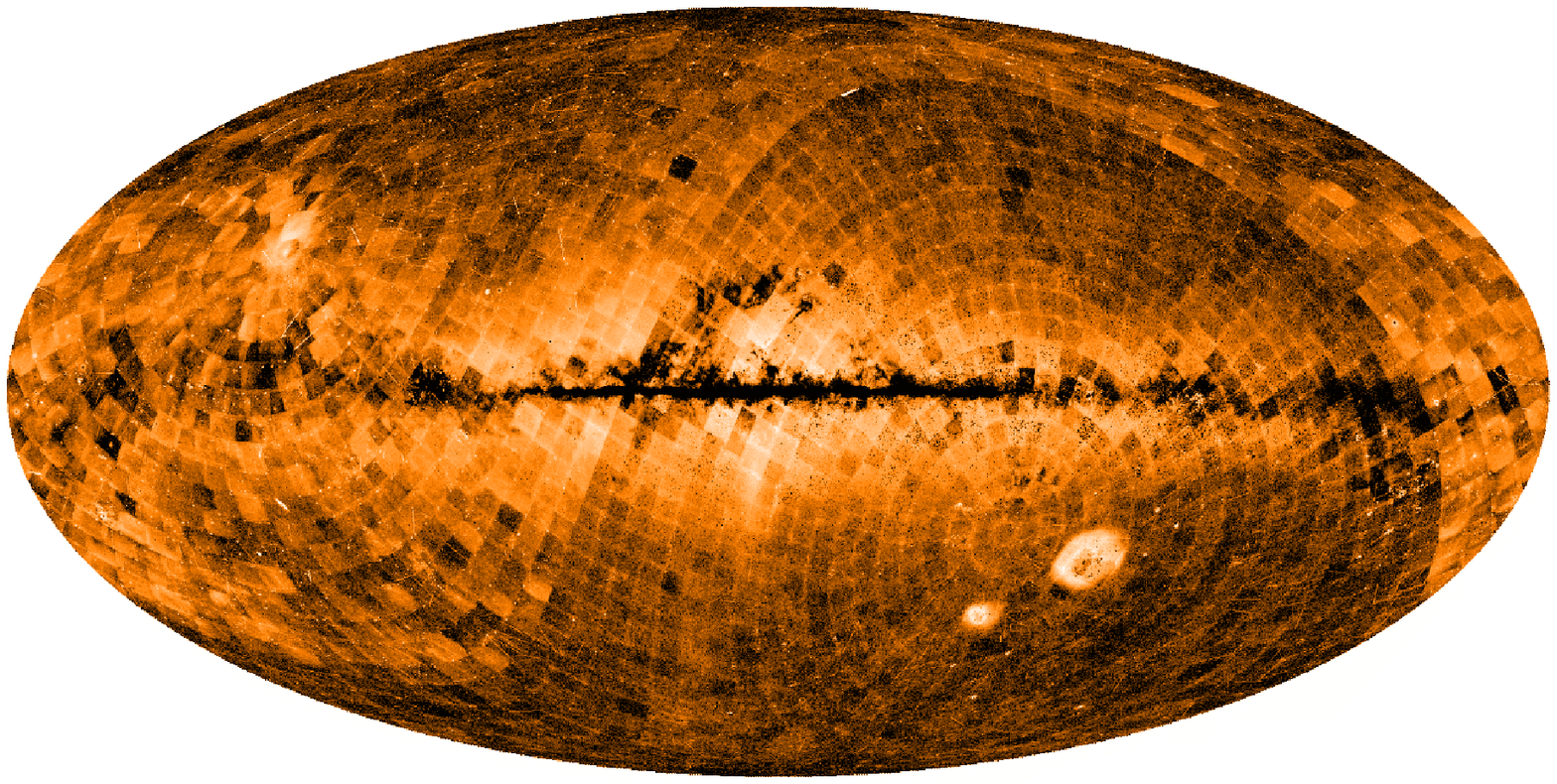}
{The sky distribution of the allsky catalogue, to an
extinction-corrected limit of $B_{\rm J}<21$, showing only
objects fainter than $R_{\rm F}=19.5$.
The Mollweide projection in Galactic coordinates
shows surface density on a logarithmic scale, from $0.1\,{\rm arcmin}^{-2}$
to $10\,{\rm arcmin}^{-2}$.
This image shows clearly the variety of artefacts that afflict a $B_{\rm J}$--only selection,
ranging from noise images in the less deep plates, to satellite trails.
}

\japfigsky{11}{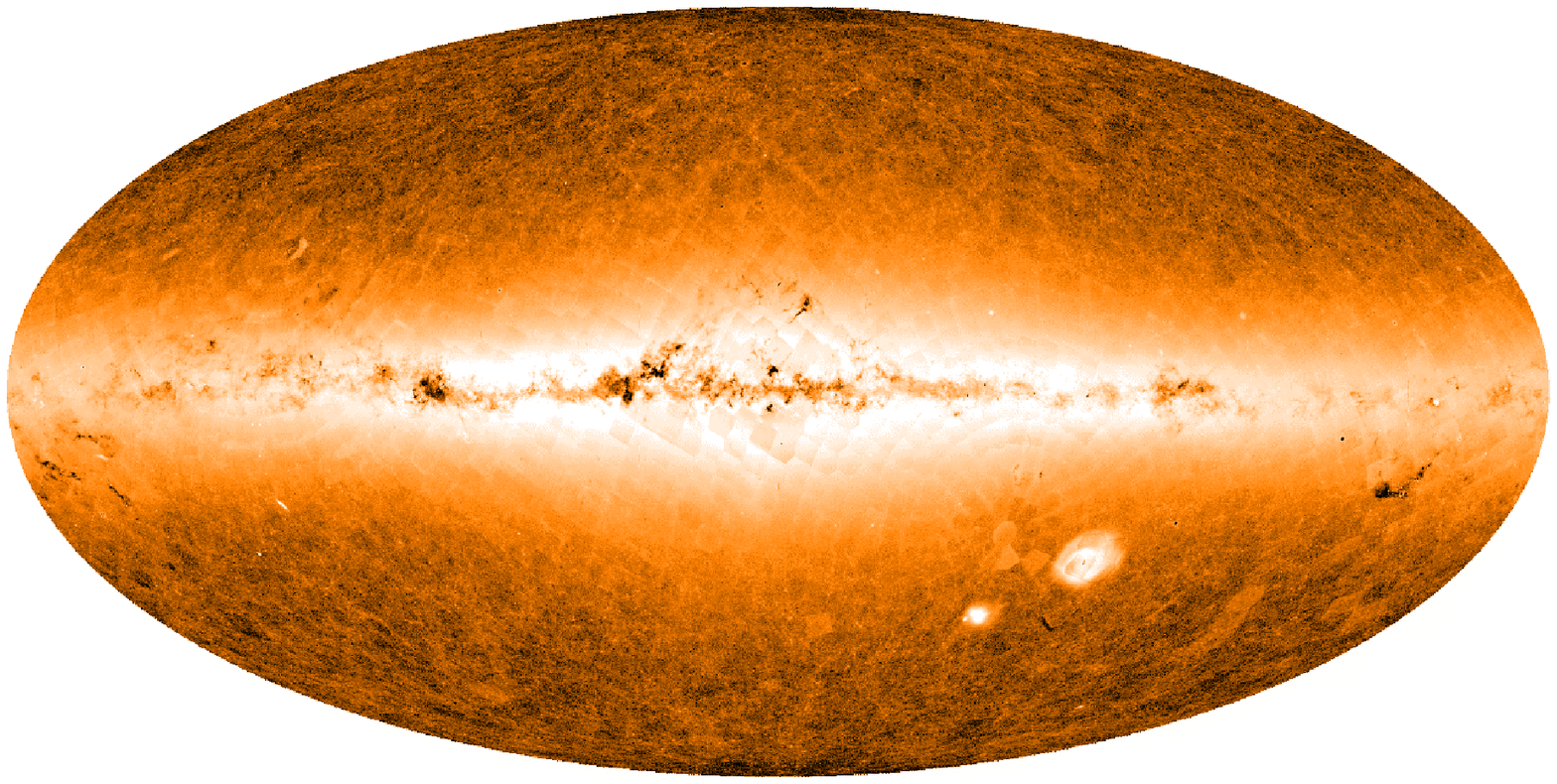}
{The sky distribution of the allsky catalogue, to an
extinction-corrected limit of $B_{\rm J}<21$, $R_{\rm F}<19.5$.
The Mollweide projection in Galactic coordinates
shows surface density on a logarithmic scale, from $0.1\,{\rm arcmin}^{-2}$
to $10\,{\rm arcmin}^{-2}$. The colours have been used to correct the southern
magnitudes to the northern system, and there is now no visible hemispheric
offset in surface density.
Compared to selection in a single waveband, the
sample is much cleaner, with the majority of artefacts removed. The main
remaining issue is spurious galaxy images in regions of high stellar
density, which cannot be removed without further information.
}

\ssec{Artefacts}

Finally, it should be pointed out that the relatively good
high-latitude performance of the SuperCOSMOS catalogue discussed 
above applies only to a sample paired up in $B_{\rm J}$
and $R_{\rm F}$. If only a single waveband is used, the
quality of the catalogue is significantly impaired.
This point is illustrated in Fig. 9, which shows the
sky density of galaxies selected to an (extinction-corrected)
limit of $B_{\rm J}<21$. Based on considerations of
$S/N$, this should be expected to be relatively clean; but
a number of artefacts are apparent. The first is the
offset in density between celestial hemispheres, owing
to the slight difference in the photometric systems as
detailed earlier. But there are also a number of other
blemishes, which can be made more obvious if we pick
out galaxies that fail to pair up with objects on
the $R_{\rm F}$ plates. We choose for this purpose a
cut at $R_{\rm F}=19.5$, and Fig. 10 plots just the
objects with $B_{\rm J}<21$, $R_{\rm F}>19.5$. This
reveals a variety of problems, ranging from large
numbers of satellite trails, together with an
inhomogeneous production of noise images in the north,
reflecting the greater variation in plate zero points
in that hemisphere (cf. Figs 2 \& 3).

In contrast with these problems, the sky distribution
of Fig. 11 is pleasingly clean. Here we require
extinction-corrected $B_{\rm J}<21$ and $R_{\rm F}<19.5$,
which also allows us to use the $B-R$ colour to correct
the magnitudes to a single system (the northern one).
This is the final dataset used as input to the
long-wavelength pairing exercises described in
Bilicki et al. (2014, 2016), and it serves as the
final extragalactic legacy of the decades of
effort that were invested in Schmidt-plate sky surveys.

\sec{CONCLUSIONS}

We have described the construction of a calibrated all-sky optical
galaxy catalogue, based on
SuperCOSMOS scans of Schmidt photographic plates from the UKST and POSS2
surveys. 

One immediate application of the catalogue has been to provide optical
photometry of the 2MASS galaxy catalogue. This allows impressively
accurate photometric redshifts to be estimated: an rms error of
$\sigma_z\simeq 0.015$, probing the galaxy distribution out to
$z\simeq 0.2$.  A preliminary version of that work was used by
Francis \& Peacock (2010a,b) in a tomographic study of the foreground
distortions of the CMB due to the Integrated Sachs-Wolfe effect.  An
in-depth study of 2MASS photometric redshifts from this dataset was
presented by Bilicki et al. (2014), accompanying a public release of the
corresponding 2MPZ catalogue. Continuing this theme of combining
the optical SuperCOSMOS data with longer-wavelength catalogues, we
have also constructed a catalogue of 18.5 million galaxies resulting
from a cross-match with WISE data (limited at Vega $W1<17$ in its 3.4-micron band;
see Wright et al. 2010), and
this is described in detail in Bilicki et al. (2016).
The photometric redshifts in this case are less precise than when 2MASS data
are included, but are still impressive ($\sigma_z/(1+z)\simeq 0.033$), and
the catalogue covers the volume out to $z=0.4$. In this way, we
hope that the SuperCOSMOS galaxy catalogue will continue to be useful
in cosmological research, while we await the happy day when 
SDSS-quality imaging becomes available over the whole sky. In principle, the
raw material for such an all-sky digital galaxy catalogue should be
available from Pan-STARRS (Kaiser et al. 2002)
and SkyMapper (Keller et al. 2007).

The SuperCOSMOS allsky galaxy catalogue described here is available at
\smallskip
{\tt http://ssa.roe.ac.uk/allSky.html}$\,$. 
\medskip
\noindent
The 2MPZ catalogue of photometric redshifts from matching SuperCOSMOS
to 2MASS and WISE (Bilicki et al. 2014) can
be found at 
\smallskip
{\tt http://ssa.roe.ac.uk/TWOMPZ.html}$\,$. 
\medskip
\noindent
The matched 
WISE$\times$SCOS catalogue described by Bilicki et al. (2016) can
be found at 
\smallskip
{\tt http://ssa.roe.ac.uk/WISExSCOS.html}$\,$.

\section*{ACKNOWLEDGEMENTS}

This research has made use of data obtained from the SuperCOSMOS
Science Archive, consisting of scanned survey plates from the UK
Schmidt Telescope and The Palomar Observatory Sky Survey
(POSS-II). This archive is prepared and hosted by the Wide Field
Astronomy Unit, Institute for Astronomy, University of Edinburgh,
which is funded by the UK Science and Technology Facilities Council.

Funding for SDSS-III has been provided by the Alfred P. Sloan Foundation, the Participating Institutions, the National Science Foundation, and the U.S. Department of Energy Office of Science. The SDSS-III web site is {\tt http://www.sdss3.org/}. SDSS-III is managed by the Astrophysical Research Consortium for the Participating Institutions of the SDSS-III Collaboration including the University of Arizona, the Brazilian Participation Group, Brookhaven National Laboratory, Carnegie Mellon University, University of Florida, the French Participation Group, the German Participation Group, Harvard University, the Instituto de Astrofisica de Canarias, the Michigan State/Notre Dame/JINA Participation Group, Johns Hopkins University, Lawrence Berkeley National Laboratory, Max Planck Institute for Astrophysics, Max Planck Institute for Extraterrestrial Physics, New Mexico State University, New York University, Ohio State University, Pennsylvania State University, University of Portsmouth, Princeton University, the Spanish Participation Group, University of Tokyo, University of Utah, Vanderbilt University, University of Virginia, University of Washington, and Yale University. 

This publication makes use of data products from the 2MASS, which is a joint project of the University of Massachusetts and the Infrared Processing and Analysis Center/California Institute of Technology, funded by the National Aeronautics and Space Administration and the National Science Foundation; the NASA/IPAC Extragalactic Database (NED), which is operated by the Jet Propulsion Laboratory, California Institute of Technology, under contract with the National Aeronautics and Space Administration and of the NASA/IPAC Infrared Science Archive, which is operated by the Jet Propulsion Laboratory, California Institute of Technology, under contract with the National Aeronautics and Space Administration. 

We thank Robert Lupton for helpful correspondence about SDSS colour
equations.  

JAP was supported by a PPARC Senior Research Fellowship
during earlier parts of this work. He is currently supported by ERC grant number 670193.
MB is supported by the Netherlands Organization for Scientific Research, NWO, through grant number 614.001.451, and through FP7 grant number 279396 from the European Research Council. 


{

\pretolerance 10000

\section*{References}

\japref Adelman-McCarthy, J.K. et al., 2008, ApJS, 175, 297
\japref Alam S. et al., 2015, ApJ Suppl., 219, 12
\japref Astier P., Pain R., 2012, CRPhy, 13, 521
\japref Aubourg \'{E} et al., 2015, PRD, 92, 123516
\japref Baldry I.K. et al., 2010, MNRAS, 404, 86
\japref Bilicki M., Jarrett T.H., Peacock J.A., Cluver M.E., Steward L., 2014, ApJ Suppl., 210, 9 
\japref Bilicki M., Peacock J.A., Jarrett T.H., Cluver M.E., et al., 2016, ApJ Suppl., in press
\japref Blair M., Gilmore G., 1982, PASP, 94, 742
\japref Colless M. et al., 2001, MNRAS, 328, 1039
\japref Colless M. et al., 2003, arXiv:astro-ph/0306581
\japref Doyle M.T. et al., 2005, MNRAS, 361, 34
\japref Efstathiou G., Sutherland W.J., Maddox S.J., 1990, Nature, 348, 705
\japref Efstathiou G. et al. (the 2dFGRS Team), 2002, MNRAS, 330, L29
\japref Francis C.L., Peacock J.A., 2010a, MNRAS, 406, 2  
\japref Francis C.L., Peacock J.A., 2010b, MNRAS, 406, 14  
\japref Fukugita M., Ichikawa T., Gunn J.E., Doi M., Shimasaku K., Schneider D.P., 1996, AJ, 111, 1748
\japref Hambly N.C. et al., 2001a, MNRAS, 326, 1279
\japref Hambly N.C., Irwin M.J., MacGillivray H.T., 2001b, MNRAS, 326, 1295
\japref Hambly N.C., Davenhall A.C., Irwin M.J., MacGillivray H.T., 2001c, MNRAS, 326, 1315
\japref Ivezi\'c {\v Z}., Smith J.A., Miknaitis G. et al.\ 2007, The Future of Photometric, Spectrophotometric and Polarimetric Standardization, 364, 165. arXiv:astro-ph/0701508
\japref Jarrett T.H., Chester T., Cutri R., Schneider S., Skrutskie M., Huchra J.P., 2000, AJ, 119, 2498
\japref Jarrett T.H., Chester T., Cutri R., Schneider S.E., Huchra J.P., 2003, AJ, 125, 525
\japref Jones D.H. et al, 2009, MNRAS, 399, 683
\japref Kaiser N. et al., 2002, Proc. SPIE, 4836, 154
\japref Keller S.C. et al., 2007, PASA, 24, 1
\japref Kov{\'a}cs A., Szapudi I., 2015, MNRAS, 448, 1305
\japref Kurcz A. et al., 2016, arXiv:1604.04229
\japref Landolt A.U., 1992, AJ, 104, 340
\japref Lupton R., 2005, {\tt https://www.sdss3.org/dr10/} \hfill\break
{\tt algorithms/sdssUBVRITransform.php}
\japref Maddox S.J., Efstathiou G., Sutherland W.J., Loveday J., 1990a, MNRAS, 243, 692
\japref Maddox S.J., Efstathiou G., Sutherland W.J., 1990b, MNRAS, 246, 433
\japref Morrissey P. et al., 2007, ApJ Suppl., 173, 682
\japref Murphy T. et al., 2010, MNRAS, 402, 2403
\japref Planck Collaboration XIII, 2015, arXiv:1502.01589
\japref Rahman M. et al. 2016, MNRAS, 457, 3912
\japref Reid I.N. et al., 1991, PASP, 103, 661 
\japref Schlafly E.F., Finkbeiner D.P. 2011, ApJ, 737, 103 
\japref Schlegel D.J., Finkbeiner D.P., Davis M., 1998, ApJ, 500, 525
\japref Stetson P., 2005, PASP, 112, 925
\japref Stetson P., 2005, PASP, 117, 563
\japref Wright E.L. et al., 2010, AJ, 140, 1868

}

\bye